\documentstyle[graphicx,lscape]{mn}

\title{X-Ray Beaming due to Resonance Scattering
in the Accretion Column of Magnetic Cataclysmic Variables}
\author [Y. Terada et al.]
        {Y. Terada,$^1$ M. Ishida,$^2$ K. Makishima,$^1$ T. Imanari,$^3$
\newauthor
        R. Fujimoto,$^4$ K. Matsuzaki,$^4$ H. Kaneda$^4$\\
        $^1$ Department of Physics, Science, The University of Tokyo\\
        $^2$ Department of Physics, Tokyo Metropolitan University\\
        $^3$ Engineering, The University of Tokyo\\
        $^4$ The Institute of Space and Astronautical Science}
\date{Accepted 2001 July 25;
      Received 2001 January 20;
      in original form 2001 May 08}
\pagerange{\pageref{firstpage}--\pageref{lastpage}}
\pubyear{2001}

\begin{document}


\maketitle
\label{firstpage}

\begin{abstract}
Extremely strong ionized Fe emission lines,
with the equivalent width reaching about 4000 eV,
were discovered with ASCA from a few Galactic compact objects,
including AX~J2315$-$0592, RX~J1802.1+1804 and AX~J1842.8$-$0423.
These objects are thought to be
binary systems containing magnetized white dwarfs (WDs).
A possible interpretation of the strong Fe-K line is 
the line-photon collimation in the WD accretion column,
due to resonance scattering of line photons.
The collimation occurs when the accretion column has a flat shape,
and the effect is augmented by the vertical velocity gradient there,
which reduces the resonant trapping of resonant photons
along the magnetic field lines.
This effect was quantitatively confirmed with Monte-Carlo simulations.
Furthermore, with ASCA observations of the polar V834 Centauri, 
this collimation effect was clearly detected
as a rotational modulation of the equivalent width of the Fe-K emission line.
Extremely strong emission lines mentioned above 
can be explained by our interpretation consistently.
Combing this effect with other X-ray information,
the geometry and plasma parameters in the accretion column were determined.

\end{abstract}


\section{Introduction}
\label{section:intro_fe}

A polar (or AM Her type object) is a binary system
consisting of a low-mass main sequence star filling its Roche lobe
and a magnetized white dwarf (WD) with $10^{7\mbox{--}8}$ G magnetic field,
which is strong enough to lock the WD spin with the orbital motion.
Matter spilling over the Roche lobe of the companion star
is captured by the magnetic field of the WD
and accretes onto its magnetic poles,
emitting hard X-rays via optically-thin thermal bremsstrahlung.

Extremely strong ionized iron emission lines have been discovered
from a few polars with the X-ray observation by ASCA.
For example, AX~J2315$-$0592 has a strong ionized Fe-K$_\alpha$ line
centered at $6.84^{+0.13}_{-0.09}$keV,
whose equivalent width (EW) reaches $900^{+300}_{-200}$ eV
(Misaki et.\ al 1996),
and RX~J1802.1$+$1804 has a strong Fe-K$_\alpha$ line
with EW $\sim 4000$ eV (Ishida et.\ al 1998).
To interpret these strong line emissions as thermal plasma emission,
the plasma metallicity needs to much exceed one solar abundance;
$\sim 2$ solar for AX~J2315$-$0592, and
$3.04 \pm 1.47$ solar for RX~J1802.1$+$1804.
Although WD binaries often exhibit highly ionized Fe-K lines,
the implied abundances are usually sub-solar,
such as $0.4^{+0.2}_{-0.1}$ solar for AM Her (Ishida et al. 1997),
$0.63\pm0.08$ solar for EX Hya (Fujimoto, Ishida 1997),
and $\sim 0.4$ solar for SS Cyg (Done, Osborne 1997).
Therefore,
we speculate that
the unusually high iron abundances of the present two WD binaries
result from some mechanisms to enhance the line EW,
rather than from high metallicities of the mass-donating stars.

We find a common feature in these two object,
which may provide a clue to the strong iron K line.
AX~J2315$-$0592 exhibits a large ($87\pm2$\%, $57\pm2\%$) orbital modulation
in the 0.7 -- 2.3 keV and 2.3 -- 6.0 keV light curves,
but almost no modulation in the 6.0 -- 10.0 keV band.
Similarly, RX~J1802.1$+$1804 exhibits
a large ($\sim$ 100\%) orbital modulation amplitude
below 0.5 keV with ROSAT observation (Greiner et al. 1998),
but the ASCA light curves are extremely flat (Ishida et al. 1998).
The lack of hard-band modulation implies that a constant fractional volume
of the accretion columns (which are optically thin to continuum X-rays) is
observed throughout the rotational phase,
and hence the inclination of the orbital plane $i$ is rather small.
Then, the soft-band modulation must be due to changes in absorption by
the pre-shock accretion flow,
indicating that we observe down onto a single pole
at the absorption maximum.
This in turn requires the magnetic co-latitude $\beta$ to be close to $i$.
In short, these systems are inferred to have $\beta \sim i \sim 0$.
In this paper, we call such a polars as ``POLE (Pole-On Line Emitter)''.

We have another example of extremely strong iron K line emitter, 
an X-ray transient source AX~J1842$-$0423,
discovered with ASCA in 1996 October
on the Galactic plane in the Scutum arm region
(Terada et al. 1999; hereafter paper I).
The most outstanding feature of this object revealed by the ASCA GIS is
the very conspicuous emission line at $6.78^{+0.10}_{-0.13}$ keV,
whose EW is extremely large at $4000^{+1000}_{-500}$ eV.
To explain this line EW, a plasma metallicity as high as
$3.0^{+4.3}_{-0.9}$ solar abundance would be required.
We found no periodicity over the period range of $62.5$ ms to a few hours,
the latter being a typical orbital period of polars.
In view of the thin-thermal spectrum
and the allowed source size of $10^{8.5 - 17.7}$ cm,
we have concluded in paper I
that AX~J1842$-$0423 is likely to another POLE,
like AX~J2315$-$0592 or RX~J1802.1$+$1804.

The face-value metallicities of the three objects are so high
that we regard these values to be unrealistic.
Instead, we consider that the iron K line EW is much enhanced by
some mechanism, which may be common to the POLEs.
To account for the strong iron K line of three POLEs,
the mechanism must account for line enhancement by a factor of 3 or more.
In this paper, we develop the possible explanation
invoking resonance scattering ($\S$\ref{section:mechanism}),
which has been proposed briefly in paper I.
In the present paper,
we carry out Monte-Carlo simulations ($\S\ref{section:monte_carlo}$)
to confirm the proposed mechanism,
and verify the effect through
ASCA observations of the polar V834 Cen ($\S\ref{section:observation}$).

\section{Line Enhancement due to Resonance Photon Beaming}
\label{section:mechanism}
\subsection{Geometrical beaming of Fe resonance line}
\label{section:mechanism_geometrical_beaming}

\begin{figure}
\centerline{\includegraphics[height=6cm]{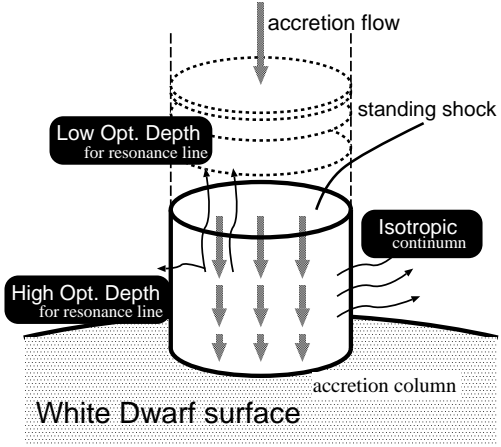}}
\caption{Schematic view of an accretion column on the WD in a polar system}
\label{fig:accretion_column}
\end{figure}

In a polar system,
a flow of matter accreting onto each magnetic pole of the WD
is highly supersonic,
so that a standing shock is formed close to the WD.
The matter is shock heated up to a temperature $k$ of a few tens keV
[Appendix \ref{section:accretion_column} equation (\ref{eq:temparature})].
The heated plasma cools by radiating Bremsstrahlung hard X-ray continuum
and line photons, as it flows down the column,
to form a hot accretion column of height $h$ and radius $r$ 
as illustrated in figure \ref{fig:accretion_column}.
We can assume that the ion temperature is equal to the electron temperature,
because the ion to electron energy transfer time scale $t_{\mbox{\tiny eq}}$
is much shorter than the cooling time scale $t_{\mbox{\tiny ff\_cool}}$
[Appendix \ref{section:accretion_column},
equations (\ref{eq:eq_timescale}) and (\ref{eq:ff_timescale})].

The typical accretion rate of polars is
$\dot{M} \sim 10^{16}$ g s$^{-1}$, $r$ is typically $5 \times 10^7$ cm,
and the velocity immediately beneath the shock front is typically
$u^{\mbox{\tiny sh}} \sim 10^8$ cm s$^{-1}$
[Appendix \ref{section:accretion_column} equation (\ref{eq:velosity})],
so that the electron density at the top of the hot accretion column
is typically $n_e^{\mbox{\tiny sh}} \sim 10^{15 \mbox{--} 16}$ cm$^{-3}$
[Appendix \ref{section:accretion_column} equation (\ref{equation:density})].
The optical depth of the column for Thomson scattering is then given by
\begin{equation}
\tau_{\mbox{\tiny T}} = 0.24 
\left( \frac{r}  {5 \times 10^7\mbox{cm}}       \right)
\left( \frac{n_e^{\mbox{\tiny sh}}}
            {7.7 \times 10^{15}\mbox{cm$^{-3}$}} \right),
\label{eq:thomson_opt}
\end{equation}
and the optical depth of free-free absorption is much smaller 
[Appendix \ref{section:accretion_column} equation (\ref{eq:opt_ff})].
Therefore, the column is optically thin
for both electron scattering and free-free absorption,
so the continuum X-rays are emitted isotropically.

On the other hand, the optical depth for the resonance scattering
is calculated from equation (\ref{eq:reso_cross_sec}) in
Appendix \ref{section:accretion_column} as
\begin{equation}
\tau_{\mbox{\tiny R}} = 36
\left( \frac{n_e} {7.7 \times 10^{15} \mbox{cm$^{-3}$}}\right)
\left( \frac{A_{\mbox{\tiny Fe}}}{4.68\times 10^{-5}}\right)
\left(\frac{r}{5 \times 10^7\mbox{cm}}\right),
\label{eq:reso_opt}
\end{equation}
at the energy of the hydrogenic iron K$_\alpha$ line,
where $A_{\mbox{\tiny Fe}}$ is the abundance of iron by number,
which is normalized to the value of one solar.
Thus, the accretion column is optically thick for resonance lines, and
the resonance line photons can only escape from positions close to
the surface of accretion column.
If the accretion column has a flat coin-shaped geometry,
and our line-of-sight is nearly pole-on to it,
we will observe the enhanced Fe-K lines.
We call this effect ``geometrical beaming''.
However, this effect can explain the Fe--K line enhancement up
to a factor of 2.0 [Appendix \ref{section:geometrical_beaming}],
which is insufficient
to explain the factor $\sim$ 3 enhancements observed from POLEs.
An additional mechanism is clearly needed.

\subsection{Additional collimation by vertical velocity gradient}
\label{section:mechanism_velocity_beaming}

\begin{figure*}
\begin{minipage}{7cm}
\includegraphics[width=7cm]{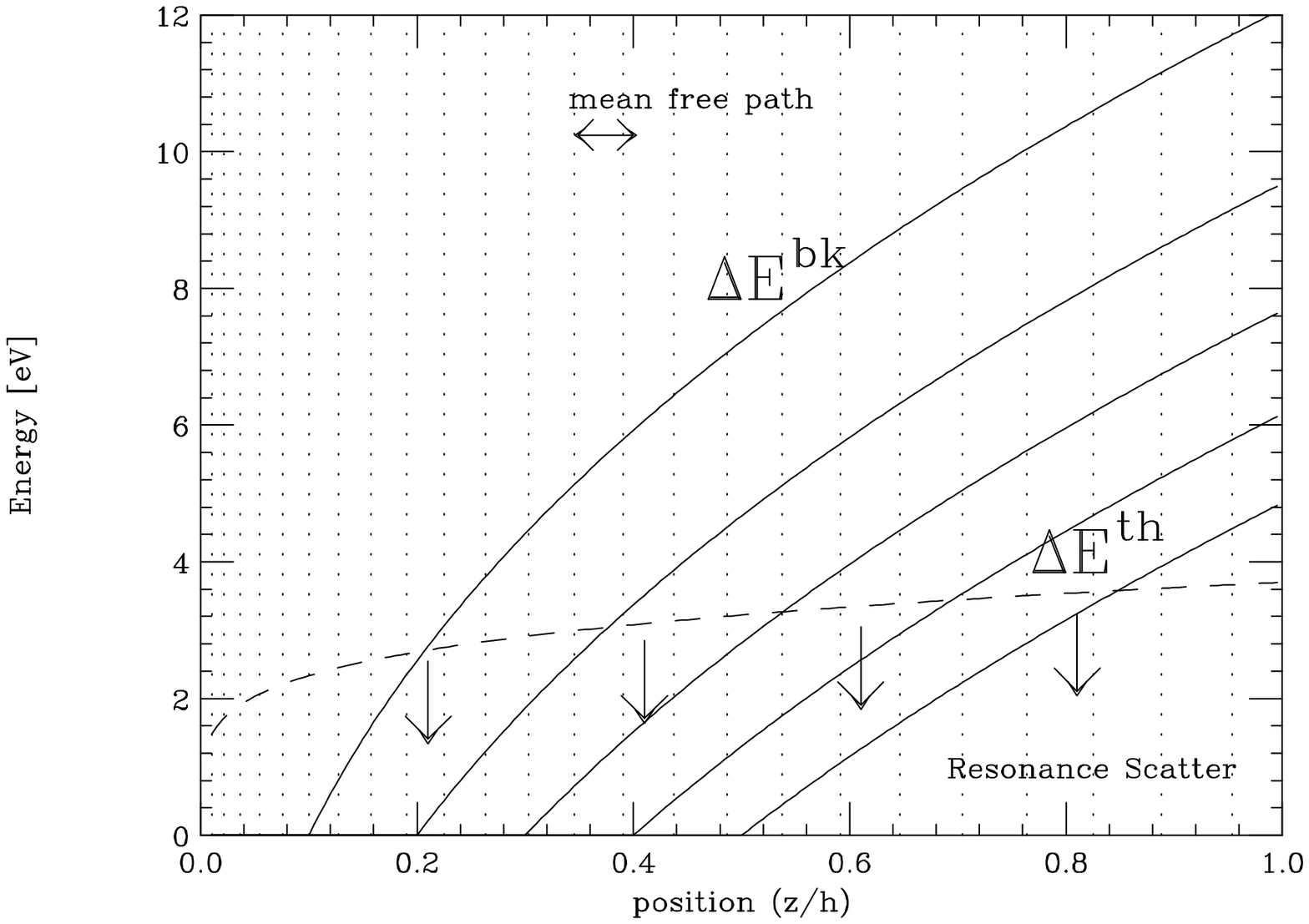}
\end{minipage}
\begin{minipage}{6.5cm}
\includegraphics[width=6.5cm]{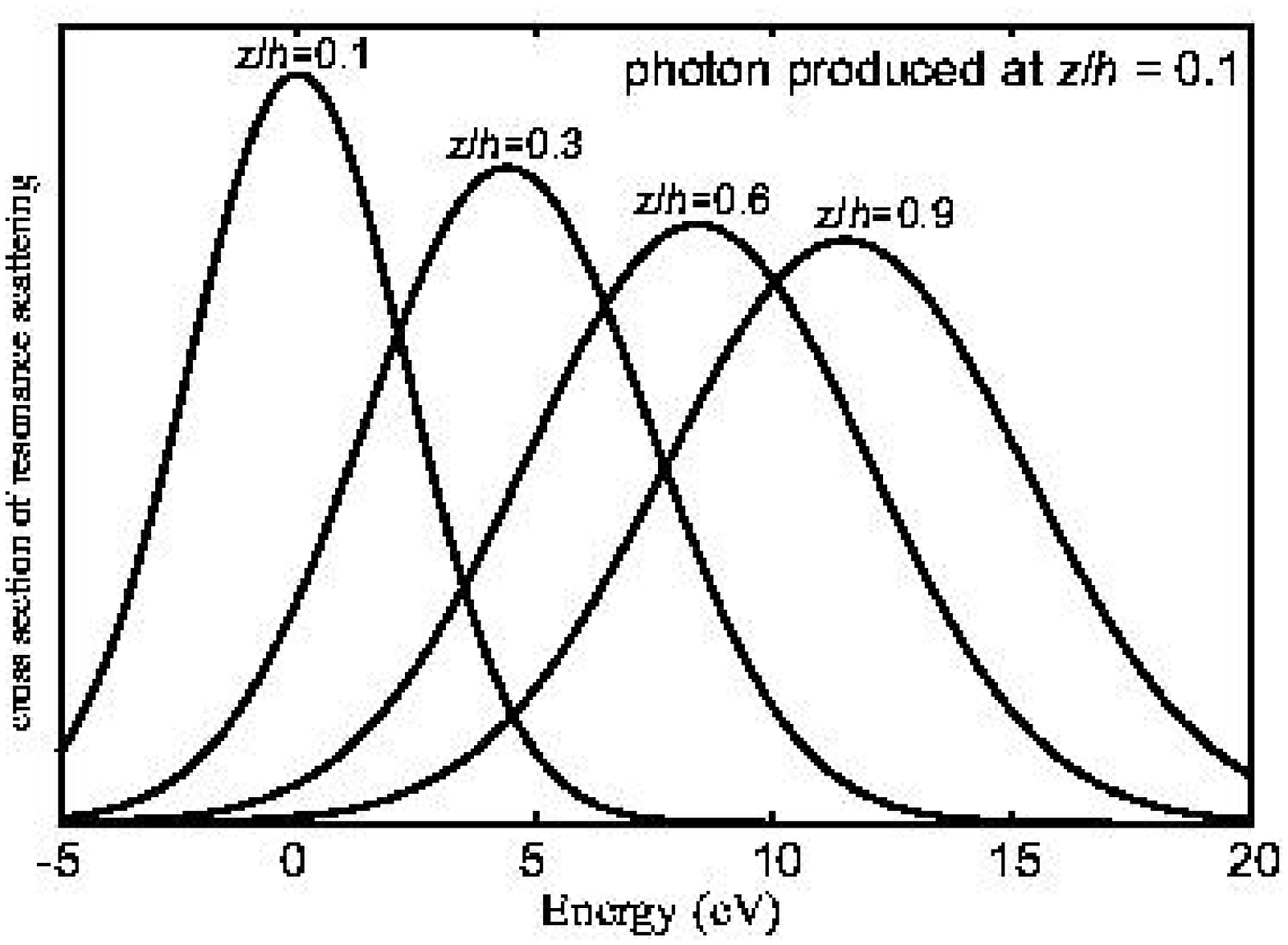}
\end{minipage}
\caption{(left) The cumulative shift of the resonance energy (solid line)
for upward-moving Fe-K line photons,
which are produced at $z = 0.1h, 0.2h, 0.3h, 0.4h$, and $0.5h$.
The thermal width of the resonant Fe-K line is given by the dashed curve.
Separations of the adjacent dotted vertical lines
specify the mean free path $l_{\mbox{\tiny R}}$ of resonant photons.
(right) Cross section of resonance scattering
for an iron K$_{\alpha}$ line photon,
which is produced at $z/h = 0.10$ and moves vertically.
The resonance centroid energy corresponds to
$\Delta E^{\mbox{\tiny bk}}$ shown in the left panel,
and the resonance energy width corresponds
to $\Delta E^{\mbox{\tiny th}}$.}
\label{fig:bulk_thermal_energy}
\end{figure*}

In the accretion column of a polar,
both temperature $kT$ and $u$ decrease, and $n_e$ increases
from the shock front toward the WD surface.
Numerically,
the vertical profiles of these quantities are calculated by Aizu (1973),
as a function of vertical distance $z$ from the WD surface,
as equation (\ref{eq:z_dependance}) in Appendix \ref{section:accretion_column}.
Because of Doppler shift caused by this strong vertical velocity gradient
in the post-shock flow, the resonance line energy changes continuously
in the vertical direction.
Let us consider, for example, that 
a line photon of rest-frame energy $E_{\mbox{\tiny 0}}$
is produced near the bottom of accretion column, $z \simeq 0.1 h$,
where the emissivity of He-like iron line photon becomes maximum,
and that this photon moves vertically by its 
mean free path of resonance scattering $l_{\mbox{\tiny R}}$
which is given as
\begin{eqnarray}
l_{\mbox{\tiny R}}
&=& \left( \sigma_{\mbox{\tiny R}} A_{\mbox{\tiny Fe}} n_e \right)^{-1}
\nonumber \\
&=& 1.3 \times 10^6
\left( \frac{z}{h} \right)^{\frac{2}{5}}
\left( \frac{n_e^{\mbox{\tiny sh}}}{7.7\times 10^{15}\mbox{cm}} \right)^{-1}
\nonumber \\ && \times 
\left( \frac{A_{\mbox{\tiny Fe}}}{4.68\times 10^{-5}}\right)
\mbox{cm}.
\end{eqnarray}
Because $h$ is given as equation (\ref{eq:column_height})
[Appendix \ref{section:accretion_column}],
$l_{\mbox{\tiny R}}$ is roughly equal to $0.07 h (\frac{z}{h})^{2/5}$.
Over this distance along the z-direction,
$u$ will change by $\Delta u = 4.0 \times 10^6$ cm s$^{-1}$
at $z \simeq 0.1h$ from equation (\ref{eq:z_dependance}),
so the resonance energy for the line photon
shifts due to Doppler effect over the same distance by
\begin{eqnarray}
\Delta E^{\mbox{\tiny bk}} &=& E_{\mbox{\tiny 0}} \frac{\Delta u}{c}
\nonumber \\
&=& 0.9 \left( \frac{E_{\mbox{\tiny 0}}}{6.695 \mbox{keV}} \right)
\left( \frac{u^{\mbox{\tiny sh}}}{ 0.9 \times 10^8 \mbox{cm $^{-3}$}} \right) 
\mbox{eV},
\label{eq:bulk_deltaE}
\end{eqnarray}
where $c$ is the light velocity.
Figure \ref{fig:bulk_thermal_energy} shows
the change of $\Delta E^{\mbox{\tiny bk}}$
for upward-moving photons produced at various depths of the accretion column.

The width $\Delta E$ of the resonance scattering
[see Appendix \ref{section:accretion_column} equation (\ref{eq:reso_cross})]
is determined
by the natural width of line photon ($\sim$ 1 eV for iron K ion)
and by the thermal Doppler broadening $\Delta E^{\mbox{\tiny th}}$.
Because the thermal velocity of ion of mass $m_{\mbox{\tiny i}}$ reaches
\begin{eqnarray}
v_{\mbox{\tiny i}} &=& \sqrt{ \frac{kT^{\mbox{\tiny sh}}}{m_{\mbox{\tiny i}}}}
\nonumber \\
&=& 2.4\times 10^7
\left( \frac{z}{h} \right)^{\frac{1}{5}}
\left( \frac{m_{\mbox{\tiny i}}}{56 m_{\mbox{\tiny H}}}\right)^{-\frac{1}{2}}
\left( \frac{ kT^{\mbox{\tiny sh}} }
{16 \mbox{keV}} \right)^{\frac{1}{2}}
\mbox{cm s$^{-1}$},
\label{eq:v_therm}
\end{eqnarray}
with $m_{\mbox{\tiny H}}$ the mass of a hydrogen atom,
the thermal Doppler width of the resonance becomes
\begin{eqnarray}
\Delta E^{\mbox{\tiny th}} (z) &=& 
E_{\mbox{\tiny 0}}
\cdot \frac{v_{\mbox{\tiny i}}}{c}
\nonumber \\
&=& 3.7 
\left( \frac{z}{h} \right)^{\frac{1}{5}}
\left( \frac{m_{\mbox{\tiny i}}}{56 m_{\mbox{\tiny H}}}\right)^{-\frac{1}{2}}
\nonumber \\ && \times
\left( \frac{E_{\mbox{\tiny 0}}}{6.695 \mbox{keV}} \right)
\left( \frac{ kT^{\mbox{\tiny sh}} }
       {16 \mbox{keV}} \right)^{\frac{1}{2}}
\mbox{eV}.
\label{eq:thermal_deltaE}
\end{eqnarray}
Here we normalized the equation to the iron atom,
$m_{\mbox{\tiny i}} = 56 m_{\mbox{\tiny H}}$.
The dashed line in figure \ref{fig:bulk_thermal_energy} shows
the position dependence of $\Delta E^{\mbox{\tiny th}}$,
below which the resonance scattering does occur.


Figure \ref{fig:bulk_thermal_energy} clearly shows that,
if a line photon gradually moves upward through repeated scattering,
its energy becomes different from the local resonance energy,
by an amount $\Delta E^{\mbox{\tiny bk}}$ [equation (\ref{eq:bulk_deltaE})]
which eventually becomes larger than the thermal width
$\Delta E^{\mbox{\tiny th}}$ [equation (\ref{eq:thermal_deltaE})].
Numerically,
this ratio for a photon produced at $z = z_0$ can be described as
\begin{eqnarray}
\frac{\Delta E^{\mbox{\tiny bk}}(z)}{\Delta E^{\mbox{\tiny th}}(z)}
&=& \frac{\Delta u}{v_{\mbox{\tiny i}}}
=   \frac{ u^{\mbox{\tiny sh}}
\left\{ \left(z/h\right)^{\frac{2}{5}} - \left(z_0/h\right)^{\frac{2}{5}}
\right\}}
{ \sqrt{ 3 \frac{\mu m_{\mbox{\tiny H}}}{m_{\mbox{\tiny i}}} }
u^{\mbox{\tiny sh}} \left(\frac{z}{h}\right)^{\frac{1}{5}} }
\nonumber \\
&=&
5.5 \left(\frac{m_{\mbox{\tiny i}}}{56 m_{\mbox{\tiny H}}}\right)^{\frac{1}{2}}
\left(\frac{\mu}{0.615}\right)
\nonumber \\ && \times 
\left\{ \left(\frac{z}{h}\right)^{\frac{1}{5}}
- \left(\frac{z_0}{h}\right)^{\frac{2}{5}}
\left(\frac{z}{h}\right)^{-\frac{1}{5}} \right\} 
\label{eq:ratio_of_bulk_therm} \\
&\simeq &
2.2 \left(\frac{m_{\mbox{\tiny i}}}{56 m_{\mbox{\tiny H}}}\right)^{\frac{1}{2}}
\left(\frac{\mu}{0.615}\right)
\nonumber \\ && \times 
\left(\frac{\delta z}{h}\right) \left(\frac{z}{h}\right)^{-\frac{4}{5}}
[ \delta z\ll h ],
\label{eq:ratio_of_bulk_therm_limit}
\end{eqnarray}
where $\delta z = z - z_0$, and 
$\mu$ is the mean molecular weight
($\mu = 0.615$ for a plasma of one-solar abundance),
and we used the relation of
\( kT^{\mbox{\tiny sh}}
= 3 \mu m_{\mbox{\tiny H}} \cdot (u^{\mbox{\tiny sh}})^2 \)
from equations (\ref{eq:temparature}) and (\ref{eq:velosity}).
Then, the photon is no longer scattered efficiently, and can escape out.
This effect does not occur in the horizontal direction
because of little velocity gradient.
As a result,
a resonant line photon produced near bottom of the accretion column
thus escapes with a higher probability
when its net displacement due to random walk is directed upward,
rather than horizontal.
In other words, iron K line photons are collimated to the vertical direction.
We call this effect ``velocity gradient beaming''.
The essence of this effect is that the mass of iron,
$m_{\mbox{\tiny i}}$ in equation (\ref{eq:ratio_of_bulk_therm}), is heavy
enough for the bulk velocity gradient to overcome the thermal line broadening.

\section{Monte-Carlo Simulation}
\label{section:monte_carlo}

The collimation effect described in section \ref{section:mechanism}
involves significant scattering processes,
which make analytic calculation difficult.
In this section, we accordingly examine the proposed effect
using Monte-Carlo simulations.

\subsection{overview of the simulation}

\begin{figure}
\includegraphics[width=7cm]{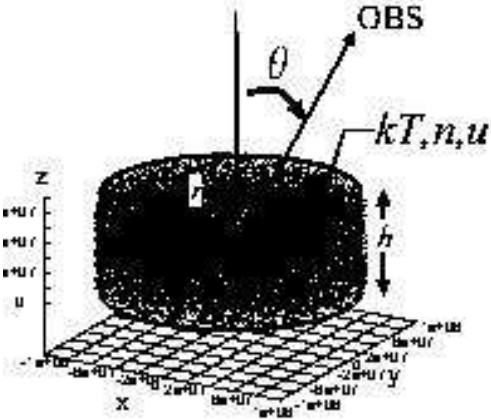}
\caption{Definition of the cylinder in our simulation.}
\label{fig:sim_cylinder}
\end{figure}

Consider a simple cylinder of height $h$ and radius $r$, 
filled with X-ray emitting plasma.
We describe the vertical dependence of $kT$, $n_e$, and $u$
in the cylinder by equation (\ref{eq:z_dependance}).
We then produce a number of Monte-Carlo iron line photons
in proportion to its emissivity,
which in turn is determined by $kT$ and $n_e$ there.
We isotropically randomize their initial direction of propagation
in the rest frame of iron nuclei.
The line energy of each iron photon is also randomized; i.e.
the average value of the energy is Doppler shifted in the frame of observers
according to the bulk velocity law [eq.\ (\ref{eq:bulk_deltaE})],
and its dispersion is determined by the local thermal motion
[eq.\ (\ref{eq:v_therm})].
We trace the propagation of each line photon with a constant step length,
which is taken to be 1/100 of the mean free path of resonance scattering 
at the bottom of the cylinder, where the temperature falls below 1 keV. 
At each step, the behavior of photon is determined by 
the calculated probabilities of resonance scattering and 
Compton scattering.
We follow the propagation of each photon until it moves outside the cylinder.

\subsection{basic condition}

\begin{figure}
\centerline{\includegraphics[width=6.5cm]{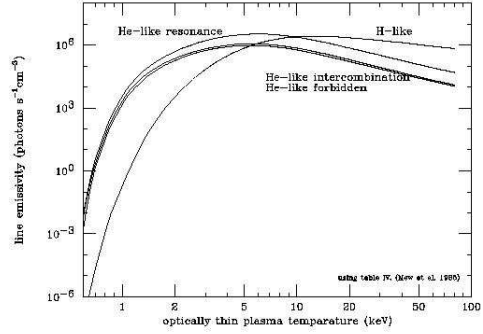}}
\caption{Temperature dependence of the Fe line emissivity (Mewe et al. 1985)
adopted in the calculation.}
\label{fig:line_pow1}
\end{figure}

The temperature dependence of iron line emissivity in optically thin plasma
has been calculated by many authors,
and we here adopt the calculation by Mewe et al. (1985)
as shown in figure \ref{fig:line_pow1}.
We consider four species of iron K line photons;
those of H-like resonance K$_{\alpha}$ line (6.965 keV),
He-like resonance  K$_{\alpha}$ line (6.698 keV),
He-like intercombination line (6.673 keV),
and He-like forbidden line (6.634 keV).
The emissivity of iron line photons per unit volume,
in erg s$^{-1}$ cm$^{-3}$, is described for each species as
\begin{equation}
P =  P'(kT) \cdot n_{\mbox{\tiny e}} n_{\mbox{\tiny Fe}} ,
\label{eq:line_power}
\end{equation}
where $ P'(kT)$ in erg s$^{-1}$ cm$^{-3}$ is the value
shown in figure \ref{fig:line_pow1}.
The iron density $n_{\mbox{\tiny Fe}}$, in cm$^{-3}$, is calculated
assuming one solar abundance.
The position dependence of iron line emissivity is
determined, through equation (\ref{eq:line_power}),
by the $z$ dependence of $kT$ and $n_{\mbox{\tiny Fe}}$
[equation (\ref{eq:z_dependance})].

\begin{figure*}
\begin{minipage}{6.5cm}
\centerline{\includegraphics[width=6.5cm]{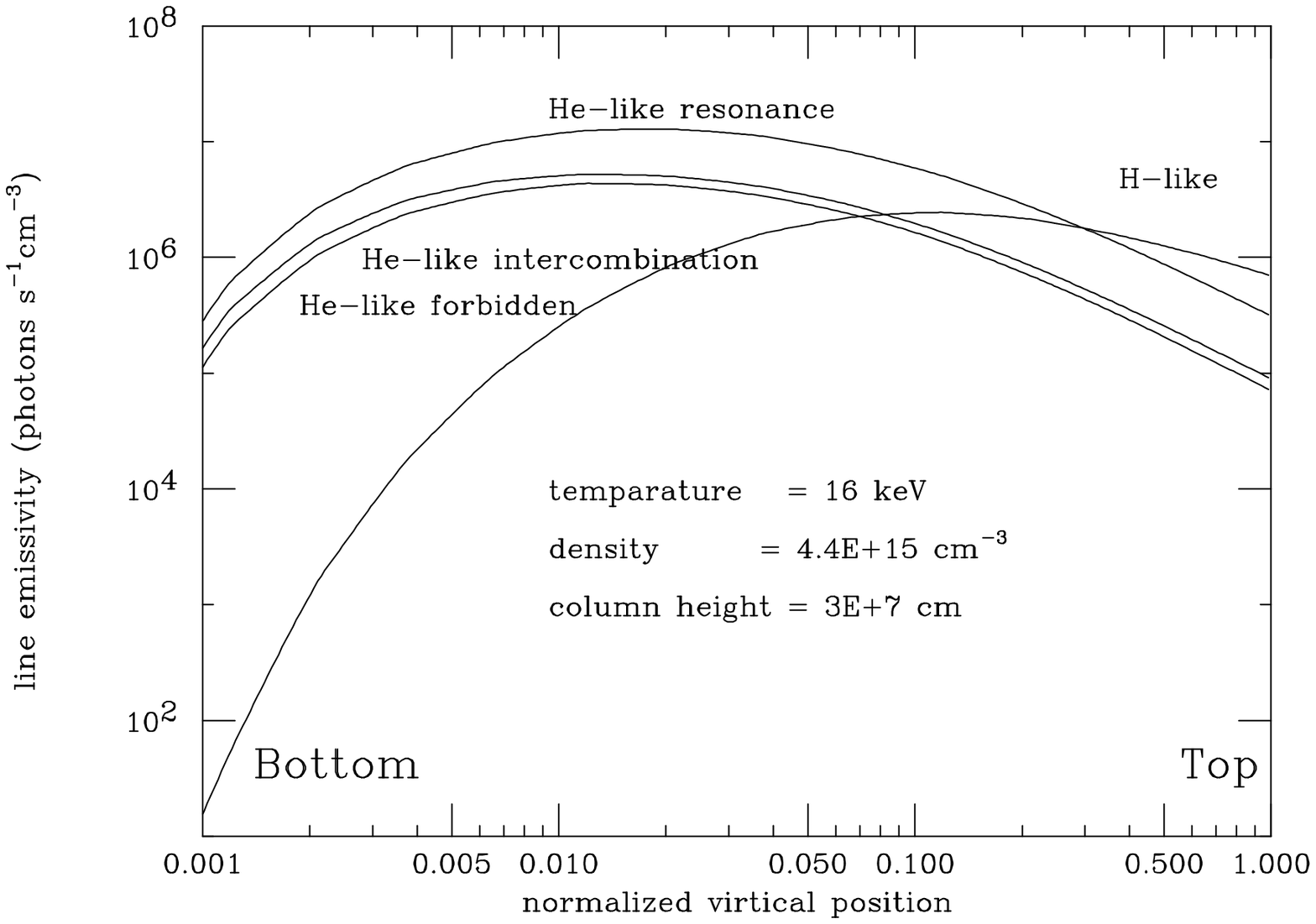}}
\end{minipage}
\begin{minipage}{6.5cm}
\centerline{\includegraphics[width=6.5cm]{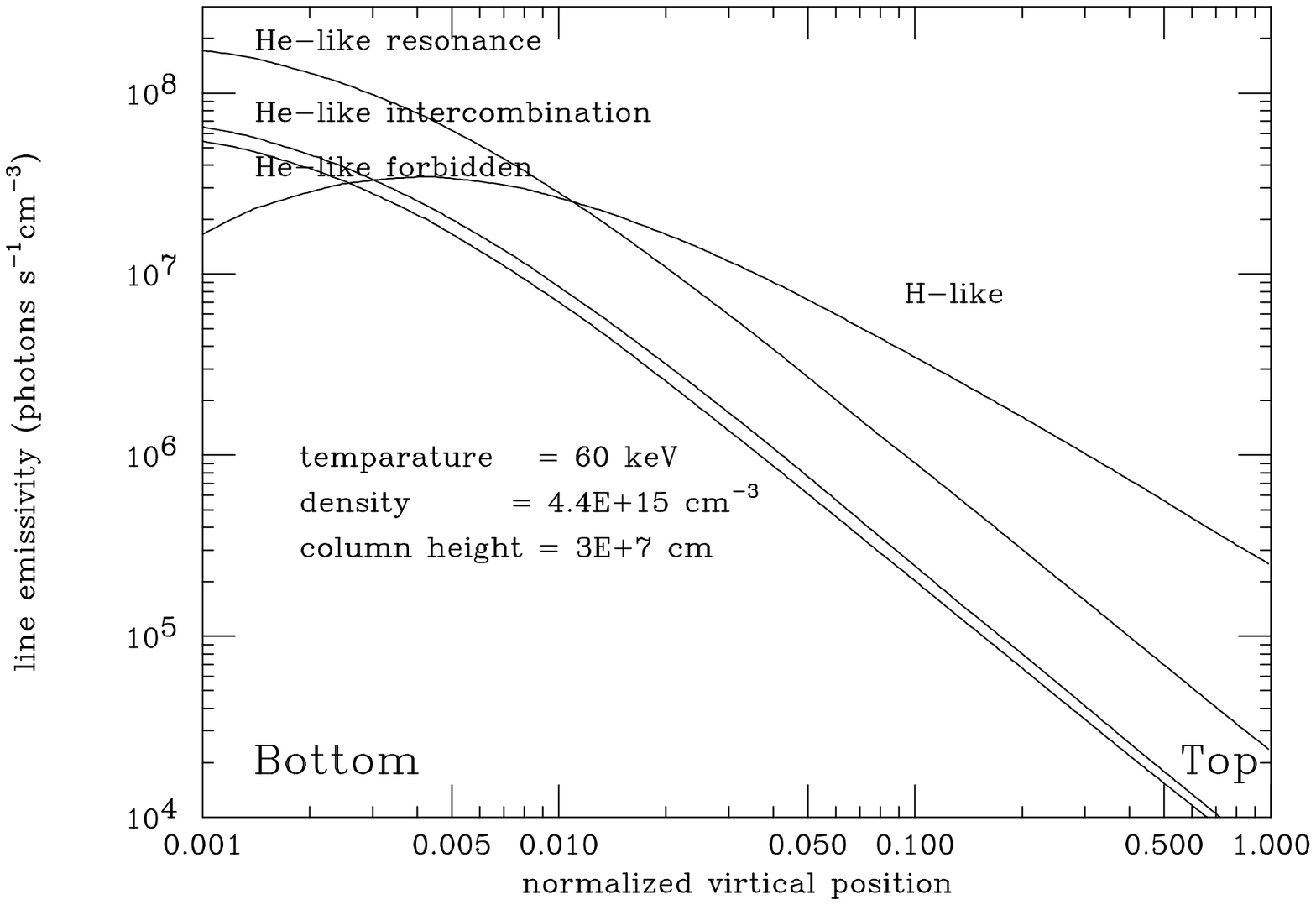}}
\end{minipage}
\caption{
The volume emissivity of Fe--K line photons in the assumed accretion column,
shown as a function of $z/h$.
The plasma temperature is $kT^{\mbox{\tiny sh}}$ = 16 keV
 (left panel), and 60 keV(right panel).}
\label{fig:column_emis}
\end{figure*}

\begin{figure}
\centerline{\includegraphics[width=6.5cm]{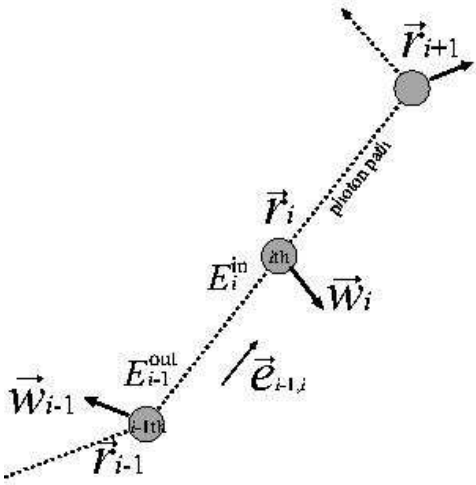}}
\caption{Schematic illustration of each step in our calculation;
$\vec{r}_i$ is the calculation site of $i$-th step,
$\vec{w}_i$ is the ion motion there in the observer's frame,
which is a sum of bulk flow motion and  random thermal motion,
and $E_i$ is the photon energy in the rest frame of $i$-th ion (See the text).}
\label{fig:monte_step}
\end{figure}

Treatment of the resonance scattering process,
taking into account both the bulk flow and the thermal motion of ions,
is a key point of the present simulation.
For each line photon being traced,
its scattering probability at the $i$-th step $\vec{r}_i$ is calculated as 
$\propto n_{\rm Fe}(\vec{r}_i) \sigma_{\mbox{\tiny RS}}(E_i^{\rm in})$,
where $n_{\rm Fe}(\vec{r}_i)$ is the local Fe-ion density 
at the position $\vec{r}_i$ 
[see Appendix \ref{section:accretion_column}
equations (\ref{equation:density}) and (\ref{eq:z_dependance})], 
$\sigma_{\mbox{\tiny RS}}$ is the cross section for the resonance scattering
given by eq.\ (\ref{eq:reso_cross})[Appendix \ref{section:accretion_column}],
and $E_i^{\rm in}$ is the Doppler-shifted energy of the incoming photon
measured in the rest frame of a representative Fe-ion at $\vec{r}_i$.
The velocity $\vec{w}_i$ of this Fe-ion, relative to the observer,
is expressed as a sum of the bulk flow velocity $u$ at $\vec{r}_i$, 
and a random thermal velocity $v$.
We specifically calculate as
\begin{equation}
E_i^{\rm in} = 
E_{i-1}^{\rm out} \left\{1+ (\vec{w}_{i-1} -\vec{w}_i) 
\cdot \vec{e_{i-1,i}} /c \right\}~,
\end{equation}
where $\vec{w}_{i-1}$ is the observer-frame velocity (bulk plus random)
of the Fe-ion that scattered the line photon last time,
$E_{i-1}^{\rm out}$ is the outgoing photon energy 
as expressed in the rest frame of the previous scatterer,
$\vec{e_{i-1,i}}$ is the unit vector along the photon propagation direction
from the $(i-1)$-th to the $i$-th scattering site
[figure \ref{fig:monte_step}].
If the scattering occurs at $\vec{r}_i$, 
we randomize the line photon energy from $E_i^{\rm in}$ to $E_i^{\rm out}$
according to the natural width,
and isotropically randomize the direction of the outgoing photon,
both in the rest frame of the present scatterer.
If, instead, the scattering does not occur at $\vec{r}_i$,
we preceed to the next step witouht changing its direction.
Thus, our calculation autonatically inclludes
both the bulk-flow Doppler effect and the thermal broadening.
However, we do not consider energy shifts by the ion recoil,
which is completely negligble.
The scattering probability for a non-resonant photon is set to 0.

In addition to the resonance scattering,
we must consider the Compton scattering process;
the energy $E$ is shifted to
\( E / \{
 1 + \frac{E}{m_{\mbox{\tiny e}} c^2}
\left( 1 - \cos \theta_{\mbox{\tiny CMP}}\right) \} \),
where $\theta_{\mbox{\tiny CMP}}$ is the Compton scattering angle.
For an iron K$_\alpha$ photon with energy $E \sim$ 6.8 keV,
a Compton scattering with $\theta_{\mbox{\tiny CMP}} \geq 10^{\circ}$
will change the photon energy beyond the resonance energy width of a few eV:
then the resonance scattering can no longer occur
after a large-angle Compton scattering.
We take this effect into account in our simulation,
using the probability distribution of $\theta_{\mbox{\tiny CMP}}$ 
by the Klein-Nishina formula,
which is almost identical to the classical formula 
for the energy of iron lines.
The differential scattering cross section and the energies of scattered photons
are calculated in the rest frame of the currently scattering electron,
so that the anisotropic effects caused by the bulk motion
 of electron is also included.
Note that we neglect the process that
the energy of a Compton-scattered continuum photon comes accidentally
into the resonance energy range,
since we do not generate continuum photons in the Monta-Carlo simulation.

\subsection{results}
\label{section:sim}

\begin{figure}
\includegraphics[width=7cm]{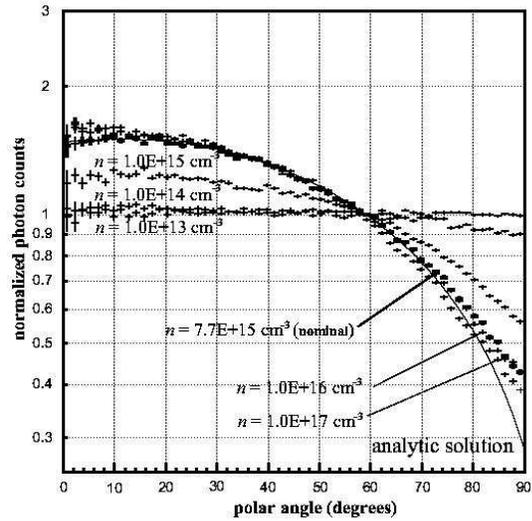}
\caption{Angular distributions of resonant line photons
emergent from a simple thin thermal plasma,
simulated by neglecting the vertical motion ($u=0$) and
neglecting the vertical gradient in $n$ and $kT$.
Abscissa is the angle $\theta$ defined in figure \ref{fig:sim_cylinder},
and ordinate is the photon flux per unit steradian
normalized to the value of an isotropic emission.
The plasma parameters are set to the nominal values;
$kT=16$ keV, $r=5\times 10^{7}$ cm, and $h=1.9\times 10^{7}$ cm.
The solid line represents
the analytic solution when only the column surface shines
 (equation \ref{eq:geometrical_profile}).
The crosses show the calculated results for various densities
as given in the figure.}
\label{fig:sim_thin_plasma}
\end{figure}

First we simulated the simplest case
wherein the plasma is hydrostatic with a single temperature
and a single density: i.e. $u$ is set to 0 and 
there is no vertical gradient in $kT$ or $n_{\mbox{\tiny e}}$.
The angular distributions of line photon flux,
calculated under this simple condition for various densities,
are shown in figure \ref{fig:sim_thin_plasma}.
The results confirm that
the photons are emitted isotropically when the plasma density is low,
but as the density increases, the geometrical beaming
becomes progressively prominent.
At $n_{\mbox{\tiny e}} = 10^{16 \mbox{--} 17}$ cm$^{-3}$,
the Monte-Carlo result agrees nicely with the analytic solution
which assumes a completely optically-thick condition;
i.e. line photons emit only from the surface of the accretion column
[equation (\ref{eq:geometrical_profile});
 see Appendix \ref{section:geometrical_beaming}].
This verifies proper performance of our Monte-Carlo simulation.

\begin{figure}
\centerline{\includegraphics[width=7cm]{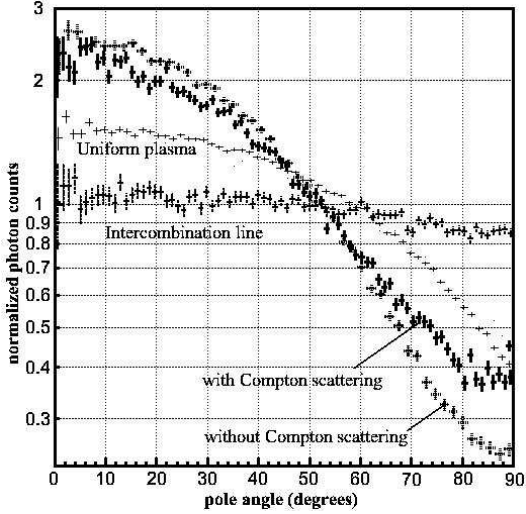}}
\caption{The same as figure \ref{fig:sim_thin_plasma},
but calculated for the He-like iron K$_{\alpha}$ lines
considering the vertical structure of the accretion column.
The plasma parameters are set to the nominal values
given in appendix \ref{section:accretion_column}:
$kT^{\mbox{\tiny sh}} = 16$ keV,
$u^{\mbox{\tiny sh}} =  0.9 \times 10^8$ cm s$^{-1}$,
$n_e^{\mbox{\tiny sh}} = 7.7 \times 10^{15}$ cm$^{-3}$,
$r = 5 \times 10^7$ cm, and $h = 1.9\times 10^7$ cm.
Thick solid crosses represent the results considering the Compton scattering,
while dashed crosses represent those neglecting the Compton process.
The dotted crosses show the profile of the intercombination line.
The thin crosses show the same profile
as presented in figure \ref{fig:sim_thin_plasma} (nominal case).}
\label{fig:sim_result_nominal}
\end{figure}

Next we have fully considered the vertical gradient
in $kT, u$, and $n_{\mbox{\tiny e}}$ [eq.\ (\ref{eq:z_dependance})].
Figure \ref{fig:sim_result_nominal} shows
the calculated angular distribution of He-like iron line
when the relevant parameters are set
to the nominal values in equations (\ref{eq:temparature}),
(\ref{eq:velosity}), and (\ref{equation:density}).
The resonance line flux is thus enhanced in the vertical direction
more strongly than in figure \ref{fig:sim_thin_plasma}.
This reconfirms the physical beaming mechanism
we proposed in $\S$ \ref{section:mechanism}.
The Compton scattering is confirmed to reduce the collimation only slightly.
Finally, the intercombination photons,
which are free from the resonance scattering,
exhibit a nearly isotropic distribution.

\begin{figure}
\includegraphics[width=8.5cm]{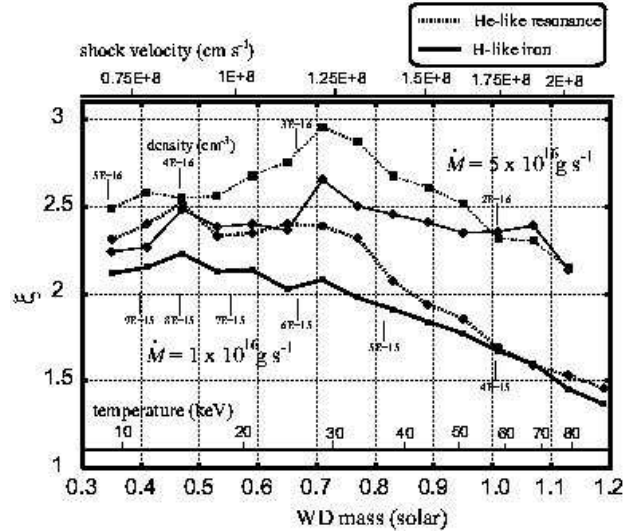}
\caption{The beaming factor $\xi_{\mbox{\tiny m}}$ with various WD mass.
The H-like iron line is represented by solid lines and
the He-like iron resonance line by dashed lines.
The baseline condition is given in the text,
and the temperatures and the densities change with WD mass and WE radius
as shown in equations (\ref{eq:temparature}) and (\ref{eq:velosity}).
The accretion rate to WD $\dot{M}$ for calculation is set to
$1 \times 10^{16}$ g s$^{-1}$ and $5 \times 10^{16}$ g s$^{-1}$.}
\label{fig:sim_result_various2}
\end{figure}

\begin{figure*}
\begin{minipage}{7.5cm}
{\bf a) column shape} \\
\includegraphics[width=7.5cm]{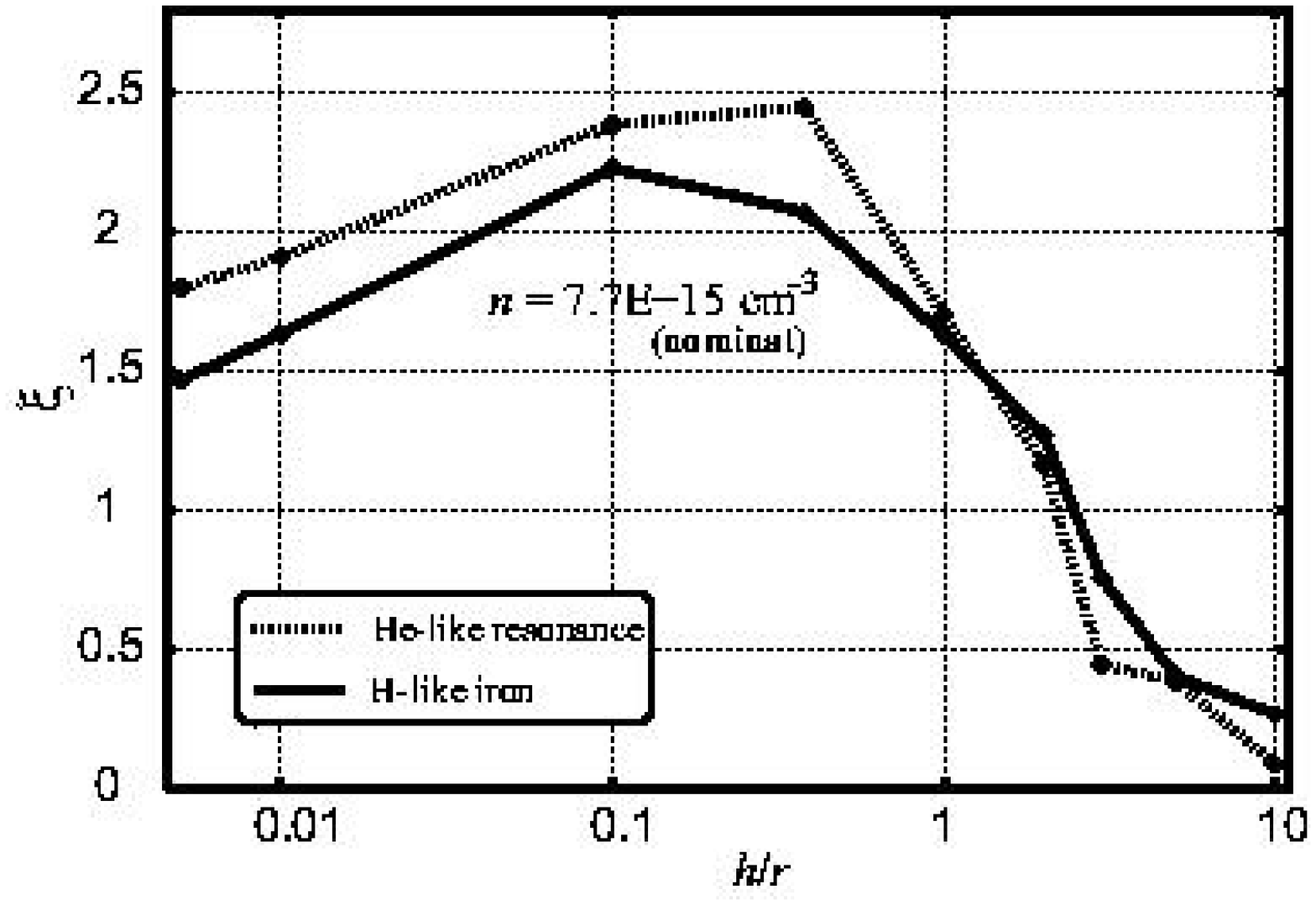}
\end{minipage}
\begin{minipage}{7.5cm}
{\bf b) electron density} \\
\includegraphics[width=7.5cm]{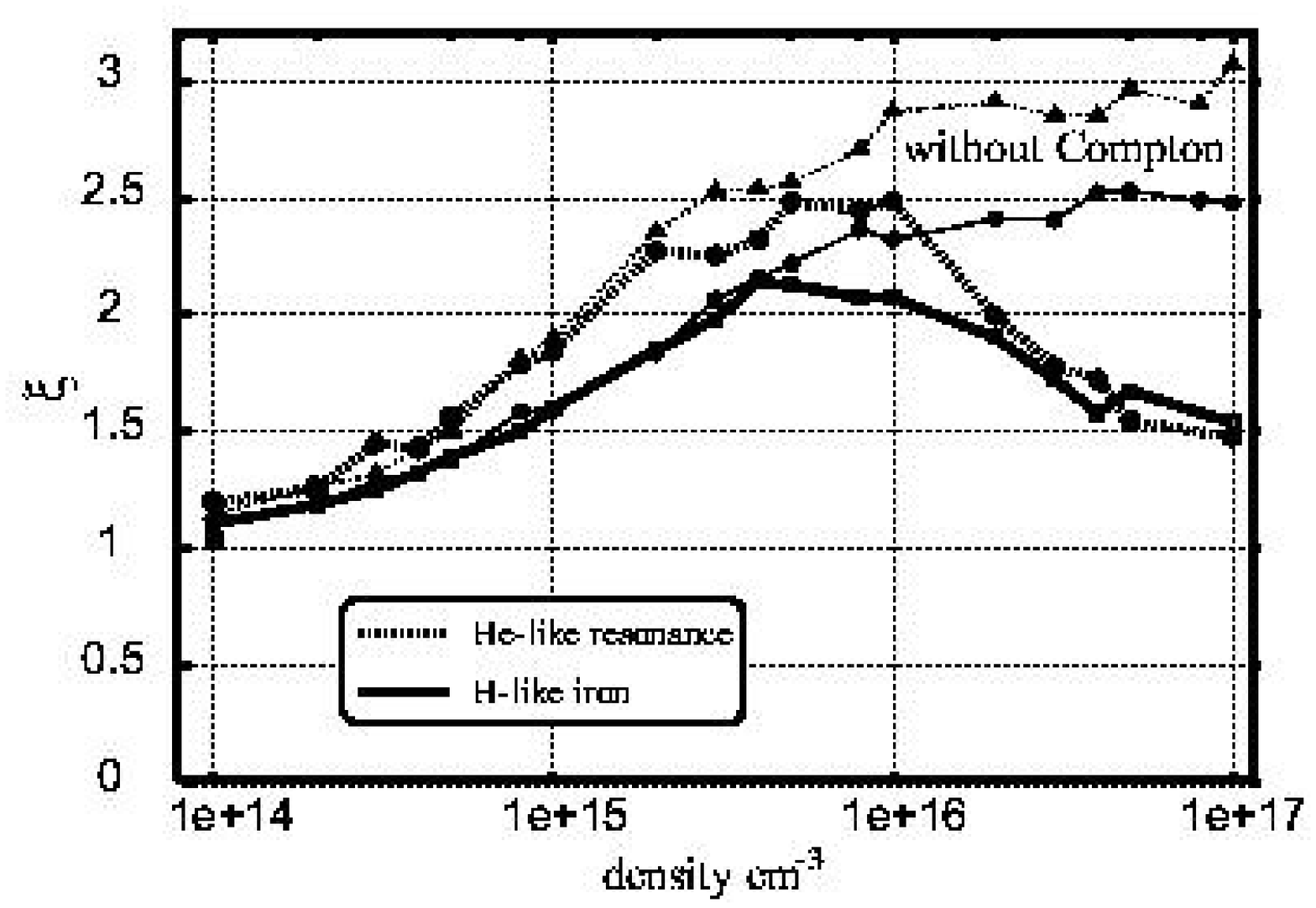}
\end{minipage}

\begin{minipage}{7.5cm}
\includegraphics[width=7.5cm]{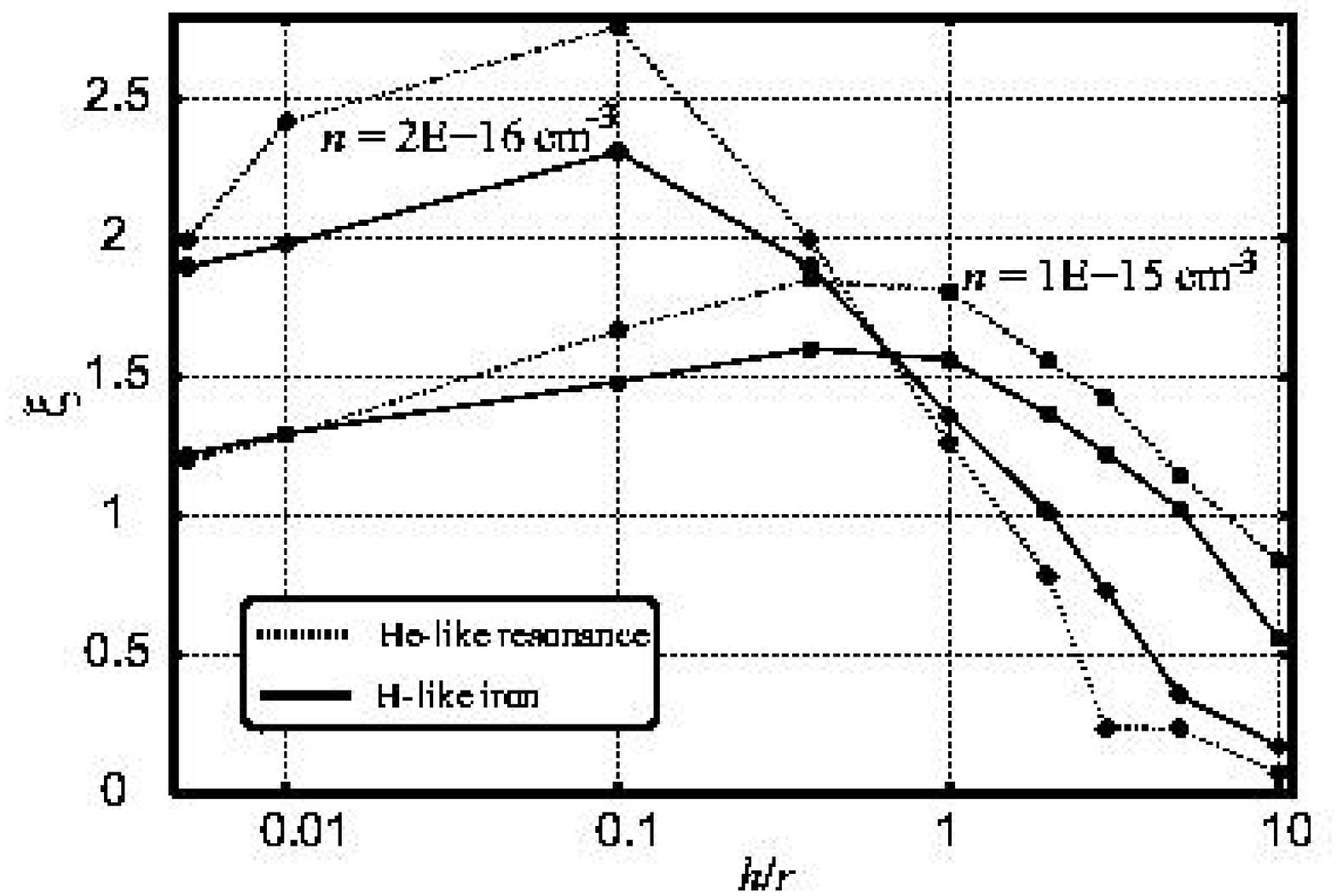}
\end{minipage}
\begin{minipage}{7.5cm}
\includegraphics[width=7.5cm]{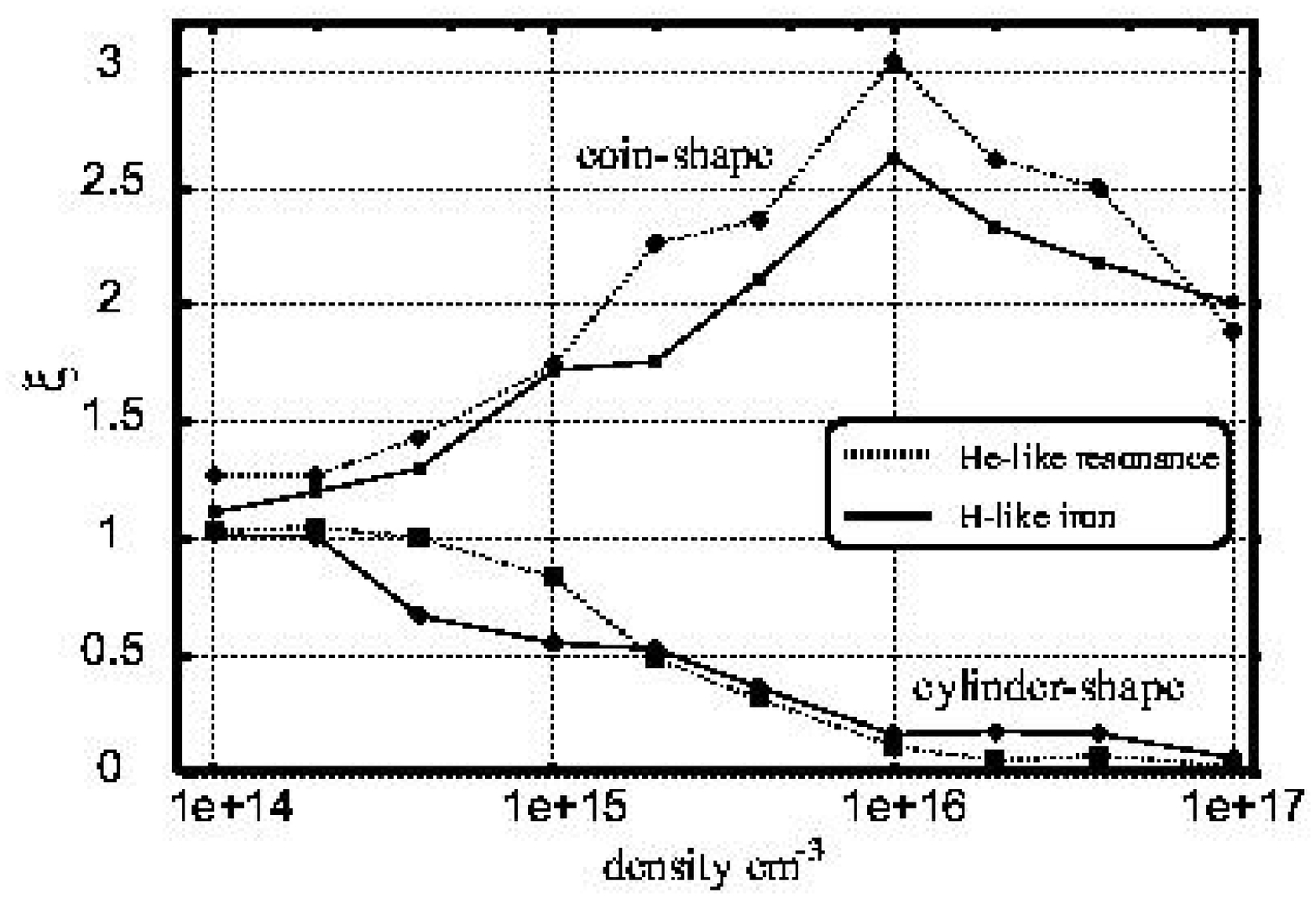}
\end{minipage}

\caption{
A summary of Monte-Carlo simulations.
The beaming factor $\xi_{\mbox{\tiny m}}$ is presented
for various column shape and density.
The H-like iron line is represented by solid lines and
the He-like iron resonance line by dashed lines.
The baseline condition of the calculation is given in the text.
{\bf a)} The shape dependence of $\xi_{\mbox{\tiny m}}$,
where the volume $r^2 h$ is fixed.
Calculation is performed for three densities as specified in the figure.
{\bf b)} The density dependence of $\xi_{\mbox{\tiny m}}$
for three different column shapes (coin shape, nominal, cylinder shape).
The coin shaped column has $r=1\times 10^{8}$ cm and $h=1\times 10^{7}$ cm,
while the cylinder shaped column has $r=1\times 10^{7}$ cm and
$h=1\times 10^{8}$ cm.
The beaming factors without Compton scattering are also shown.}
\label{fig:sim_result_various}
\end{figure*}

We repeated the Monte-Carlo simulations by changing
$h/r$, $n^{\mbox{\tiny sh}}$, $kT$, and $u^{\mbox{\tiny sh}}$,
around their baseline values of 
$kT = 16.0$ keV, $u^{\mbox{\tiny sh}} = 0.9 \times 10^8$ cm s$^{-1}$,
$n_e^{\mbox{\tiny sh}} = 7.7 \times 10^{15}$ cm$^{-3}$,
$r = 5 \times 10^{7}$ cm, and $h = 1.9\times 10^7$ cm
(Appendix \ref{section:accretion_column}).
Figure \ref{fig:sim_result_various} summarizes the obtained results
in terms of the beaming factor
\begin{equation}
\xi_{\mbox{\tiny m}} \equiv
\frac{f(0)}{\int_{0}^{\frac{\pi}{2}} f(\theta) d\cos\theta}
\end{equation}
where $f(\theta)$ is the angular distribution of line photon flux,
such as is shown in figure \ref{fig:sim_result_nominal}.
The beaming factor $\xi_{\mbox{\tiny m}}$ increases
as $h/r$ decreases (i.e. coin shaped column),
or density increases (figures \ref{fig:sim_result_various}a and b).
However, when the density exceeds $\sim 10^{16}$ cm $^{-3}$,
the beaming effect diminishes again,
because of large-angle Compton scattering.
This inference is achieved by comparing results
with and without Compton process (figure \ref{fig:sim_result_various}b).
The WD mass dependence of $\xi_{\mbox{\tiny m}}$ is small
(figure \ref{fig:sim_result_various2}):
it increases slightly with mass increases
because shock velocity increases with deeper gravity potential,
and it starts decreasing because density decreases.
These results show clearly that
the strong collimation of He-like iron K$_{\alpha}$ photons,
with $\xi_{\mbox{\tiny m}} \geq 2$, is possible under reasonable conditions.

\section{An Observational Approach}
\label{section:observation}

In order to experimentally verify our interpretation,
it is necessary to measure the equivalent width of the resonant Fe--K lines
as a function of viewing angle.
For that purpose, we may utilize a polar of which our line-of-sight
relative to the magnetic axis changes from $\sim 0^{\circ}$ (pole-on)
to $\sim 90^{\circ}$ (side-on),
as the WD rotates.
Among the polars with well determined system geometry
(via optical, ultraviolet and infrared observations),
V834 Cen is particularly suited:
its orbital plane is inclined to our line-of-sight
by $i=45\pm9^\circ$,
and its magnetic co-latitude is $\beta=25\pm5^\circ$ (Cropper 1990).
As a result, our line-of-sight to the accretion column changes 
from $20^{\circ}$ to $70^{\circ}$.

\subsection{observations of V834 Cen with ASCA}
ASCA has four X-ray Telescopes (XRT:Selemitsos et.al 1995), and
its common focal plane is equipped with two Gas Imaging Spectrometers 
(GIS: Ohashi et al.\ 1996; Makishima et al.\ 1996) and 
two Solid-state Imaging Spectrometers
(SIS: Burke et al. 1991; Yamashita et al. 1997).
The ASCA observation of V834 Cen was carried out
for about 20 ksec from 1994 March 3.63 to 1994 March 4.13 (UT), and
about 60 ksec from 1999 February 9.93 to 1999 February 10.72 (UT).
In these observations, the GIS was operated in PH-nominal mode,
which yields 0.7--10.0 keV X-ray spectra in 1024 channels, and
the SIS was operated in 1-CCD FAINT mode, which produces
0.4--10.0 keV spectra in 4096 channels.
The target was detected with
a mean count rate of 0.171 c s$^{-1}$ per GIS detector and
0.259 c s$^{-1}$ par SIS detector in 1994.
The corresponding count rates were
0.191 c s$^{-1}$ and 0.253 c s$^{-1}$ in 1999.

For extracting the source photons,
we accumulated the GIS and SIS events
within a circle of radius $4'\hspace{-4pt}.5$ centered on V834 Cen,
employing the following data-selection criteria.
We discarded the data during the ASCA passing
through the South Atlantic Anomaly.
We rejected the events acquired when the field of view (FOV) of ASCA
was within 5$^\circ$ of the Earth's rim.
Furthermore for the SIS, we discarded
the data acquired when the FOV is within 10$^\circ$ of the bright Earth rim
and those acquired near the day-night-transition of the spacecraft.

\subsection{Light curves}
\begin{figure*}
\centerline{\includegraphics[width=15cm]{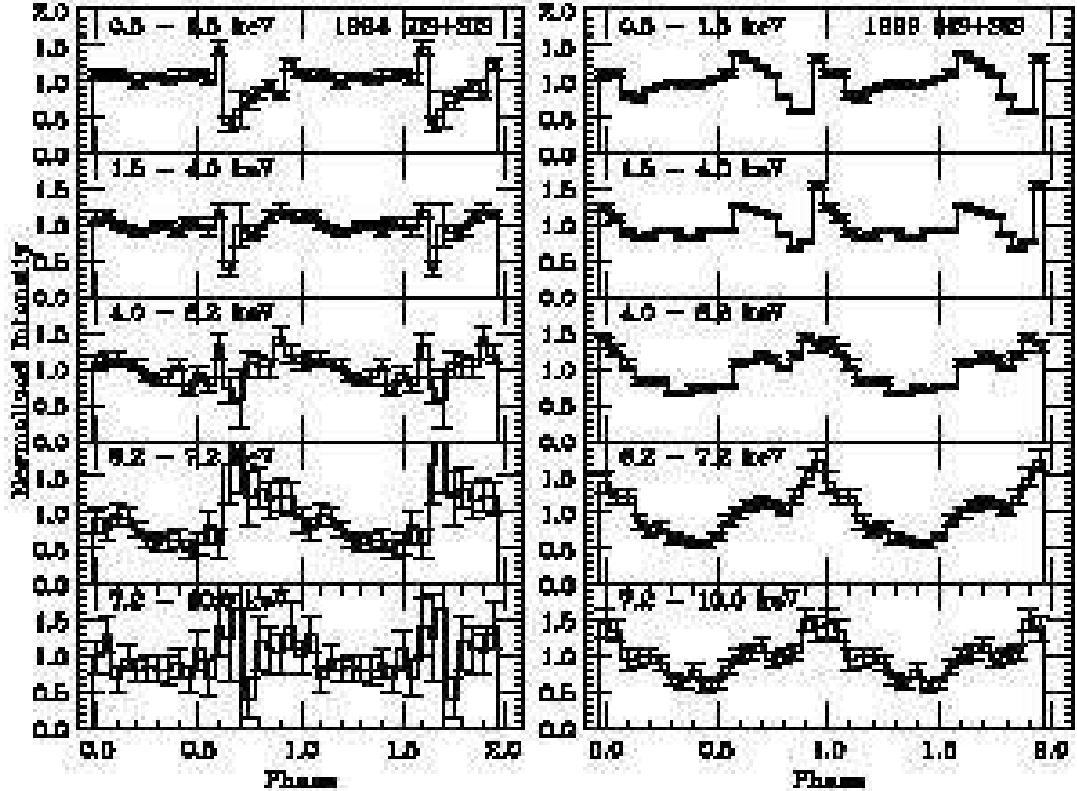}}
\caption{Energy resolved light curves of V834 Cen obtained with ASCA,
folded at 1.69194 hr.
Phase $\phi=0$ corresponds to HJD 2445048.9500,
which is common to the 1994 and 1999 light curves.
Each light curves is shown for two cycles.
Left panel shows the observation in 1994, right panel in 1999.
}
\label{fig:folded_lc}
\end{figure*}

Figure \ref{fig:folded_lc} shows the energy resolved light curve of V834 Cen
obtained with ASCA SIS$+$GIS
folded by its rotational period, $1.69194$ hr (Schwope et al. 1993).
The phase is coherent between the two light curves.
The pole-on phase of V834 Cen is determined by
the optical photometry and polarimetry 
as shown in table 1 of Bailey et al.\ (1983),
which corresponds to a phase $\phi \sim 0.65 \mbox{--} 0.85$
in our X-ray light curve.
We can recognize small dips
in light curves of the softer two energy bands in figure \ref{fig:folded_lc}
at $\phi \sim 0.65$ in 1994 and $\sim 0.86$ in 1999.
These dips are thought to arise from photoelectric absorption
by the pre-shock matter on the accretion column.
Therefore, the pole-on phase is consistent between 
the optical and X-ray datasets.
The folded light curve in the iron line energy band (6.2 -- 7.2 keV)
exhibits a hump at or near this pole-on phase,
suggesting that the proposed line photon enhancement is
indeed taking place in this system.
However, detailed examination of the modulation of iron line emission
needs phase-resolved spectroscopy, performed in the next subsection.

\subsection{spectral analysis}
\begin{figure*}
\begin{minipage}{7cm}
{\bf Pole on spectrum}\\

\centerline{
\includegraphics[height=4.6cm]{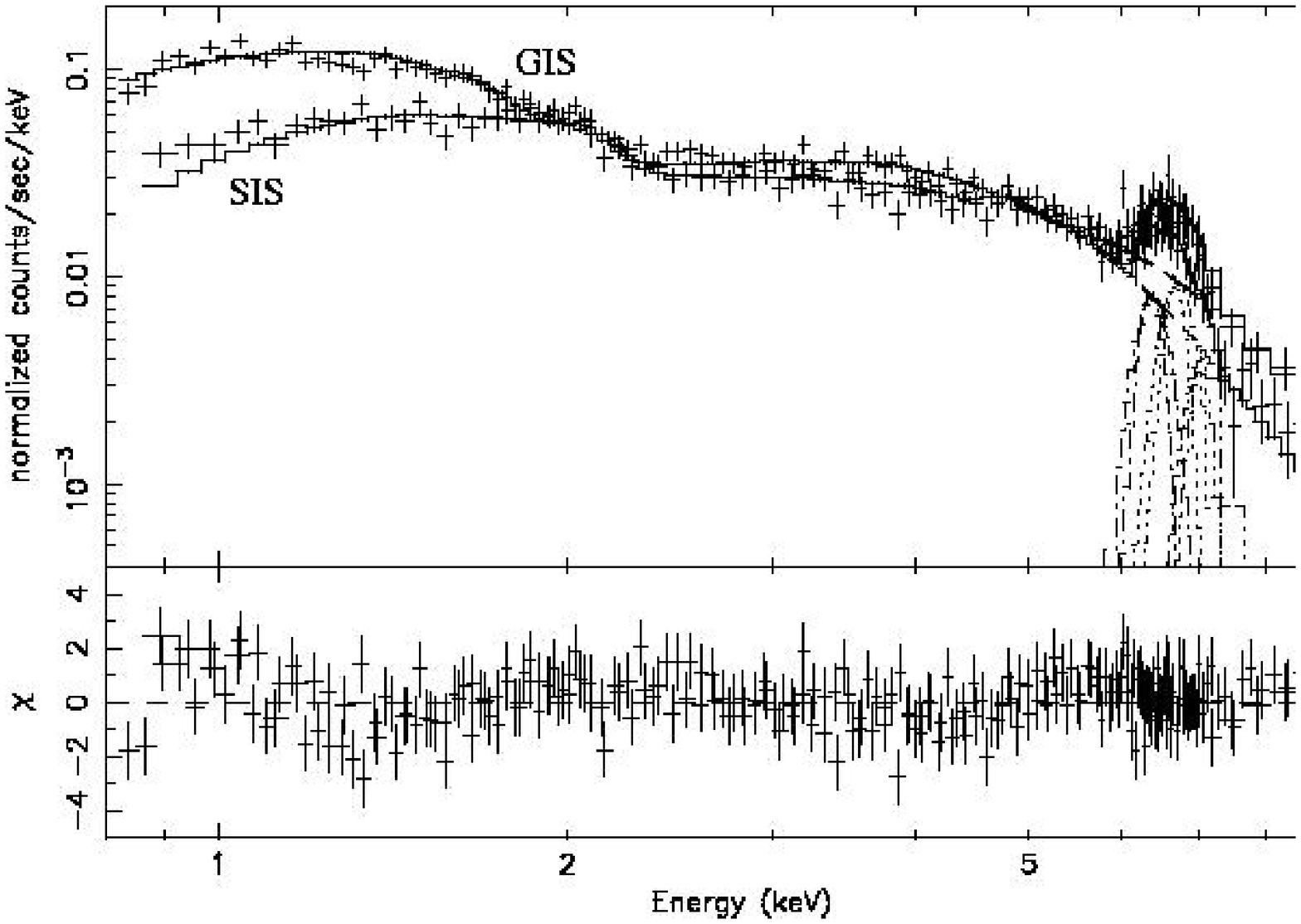}}
\end{minipage}
\hspace*{-10pt}
\begin{minipage}{7cm}
{\bf Side on spectrum}\\

\centerline{
\includegraphics[height=4.6cm]{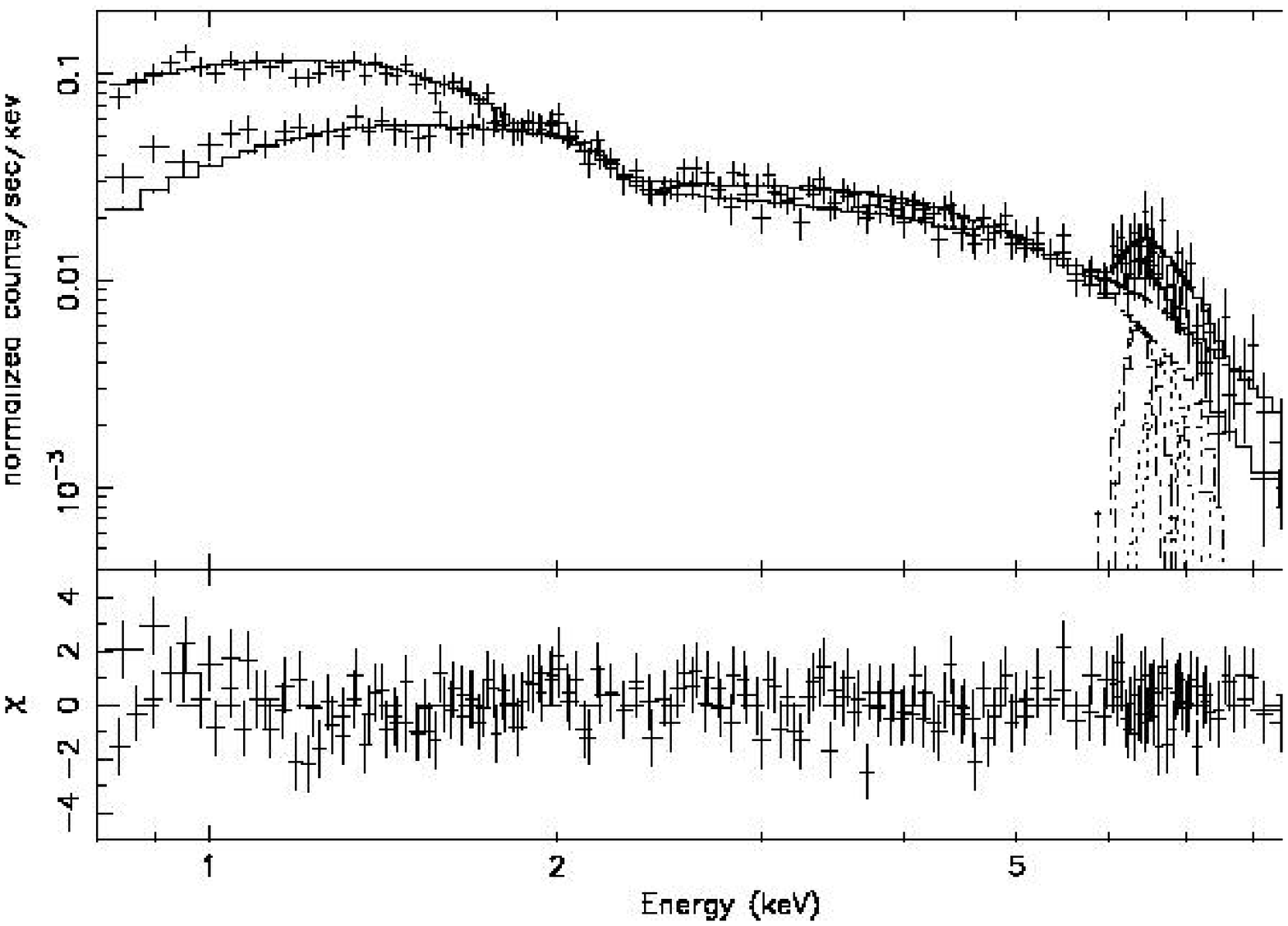}}
\end{minipage}
\caption{ASCA X-ray spectra of V834 Cen taken in 1999,
shown without removing the instrumental responses.
Left panel shows the spectra in the pole-on phase ($\phi = 0.73 \pm 0.25$),
and right panel those in the side-on phase ($\phi = 0.23 \pm 0.25$).
The GIS and SIS data are shown with crosses.
The solid crosses show 
the best-fit model consisting of single temperature bremsstrahlung continuum
with partially covered absorption and three narrow Gaussians (see text).
The best-fit parameters are shown in table \ref{tbl:fit_continuum}
Model 4 and in table \ref{tbl:fit_line} (full band fitting).
Lower panels show the fit residuals.}
\label{fig:full_band_spec_best}
\end{figure*}

Because the folded light curves (figure \ref{fig:folded_lc})
have different shapes between the two observations,
and because spectral information of GIS-3 in 1994 was degraded
by a temporary malfunctioning in the GIS onboard electronics,
we use only the 1999 data for spectral analysis.
We have accumulated the GIS (GIS2 + GIS3) and SIS (SIS0 + SIS1) data
over the pole-on phase ($\phi = 0.73 \pm 0.25$) and
side-on phase ($\phi = 0.23 \pm 0.25$) separately.
We subtracted the background spectrum,
prepared by using the blank sky data of the GIS and SIS.
The spectra, thus derived and shown in figure \ref{fig:full_band_spec_best},
exhibit an absorbed continuum with
strong iron K$_{\alpha}$ emission lines over the 6.0--7.2 keV bandpass.

\subsubsection{Continuum spectra}

\begin{figure*}
\begin{minipage}{7cm}
{\bf single $kT$, single $N_{\mbox{\tiny H}}$}\\
\vspace*{-1.6cm}

\centerline{\includegraphics[height=8cm,angle=-90]{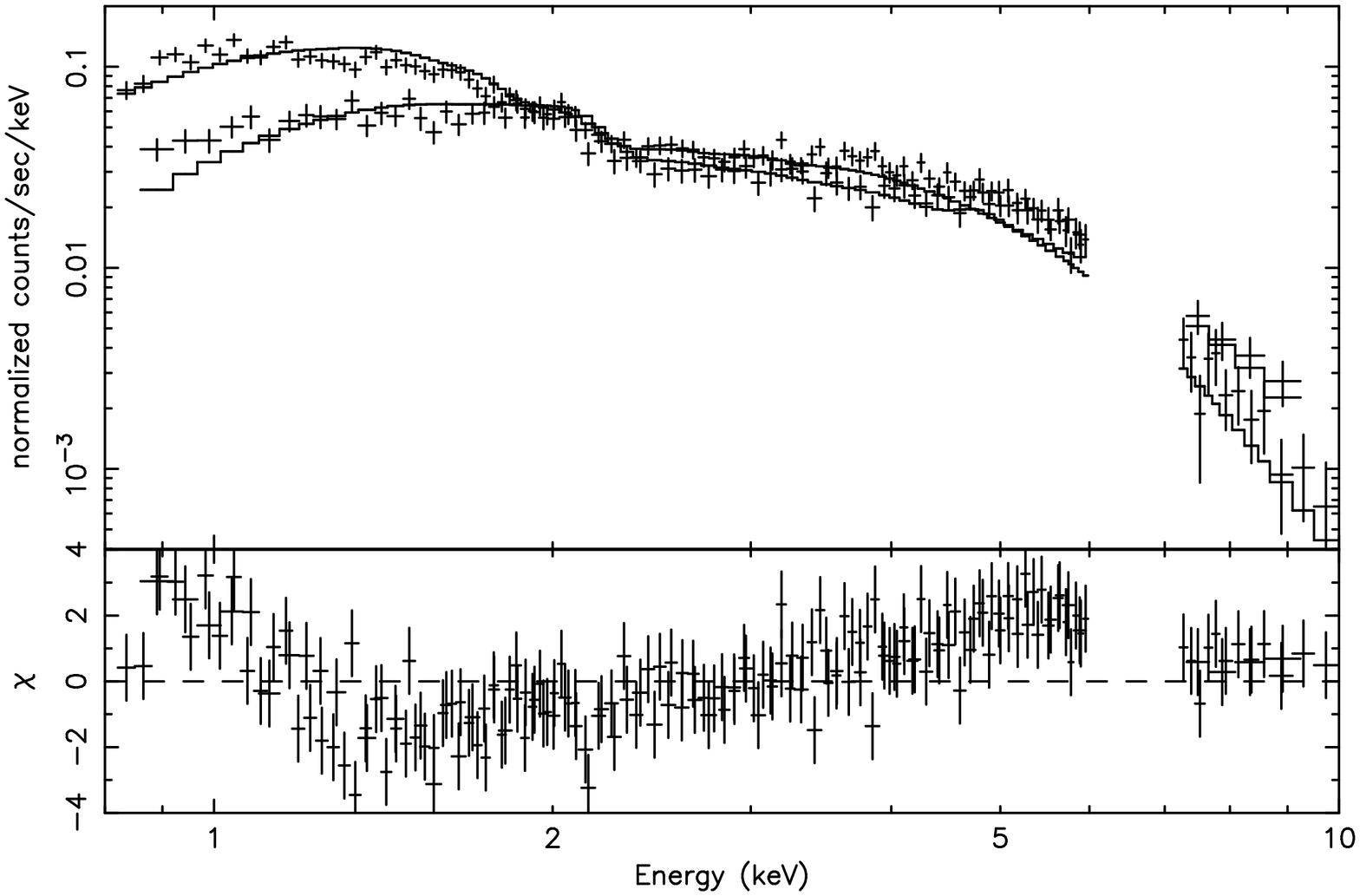}}
\end{minipage}
\hspace*{-10pt}
\begin{minipage}{7cm}
{\bf multi $kT$, single $N_{\mbox{\tiny H}}$}\\
\vspace*{-1.6cm}

\centerline{\includegraphics[height=8cm,angle=-90]{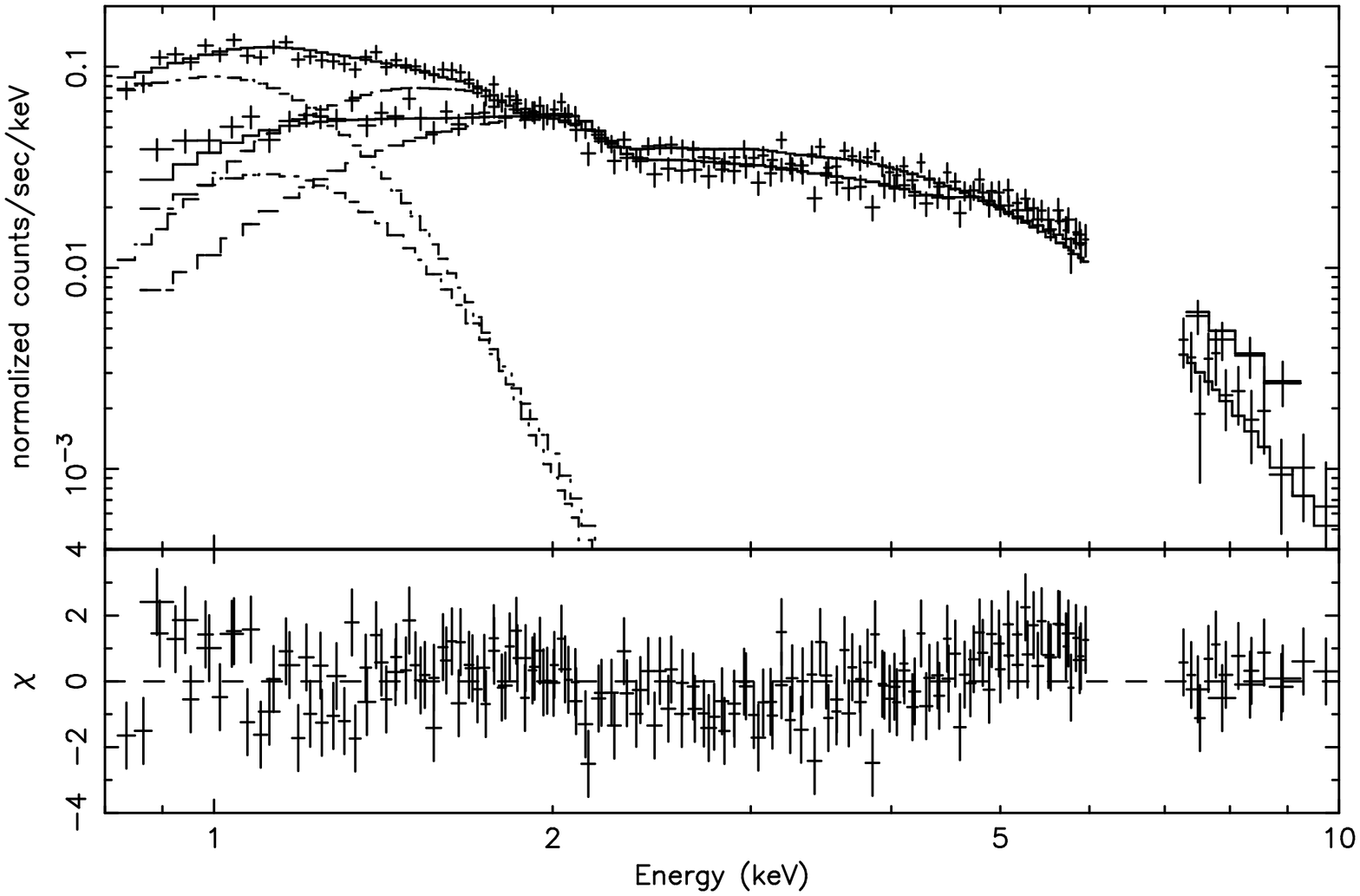}}
\end{minipage}
\caption{The same spectra as in figure \ref{fig:full_band_spec_best} left
(pole-on).
Left panel shows the best-fit model
of single temperature bremsstrahlung absorbed by a single column density.
The best-fit parameters are shown in table \ref{tbl:fit_continuum} Model 1.
Right panel shows the best-fit model
of double temperature bremsstrahlung absorbed by a single column density,
corresponding to table \ref{tbl:fit_continuum} Model 3 (free $kT$).}
\label{fig:full_spec_discarded_fit}
\end{figure*}

\begin{figure*}
\begin{minipage}{4.4cm}
{\bf a. single narrow}\\
\vspace*{-1.5cm}
\includegraphics[height=5.8cm,width=4.4cm,angle=-90]{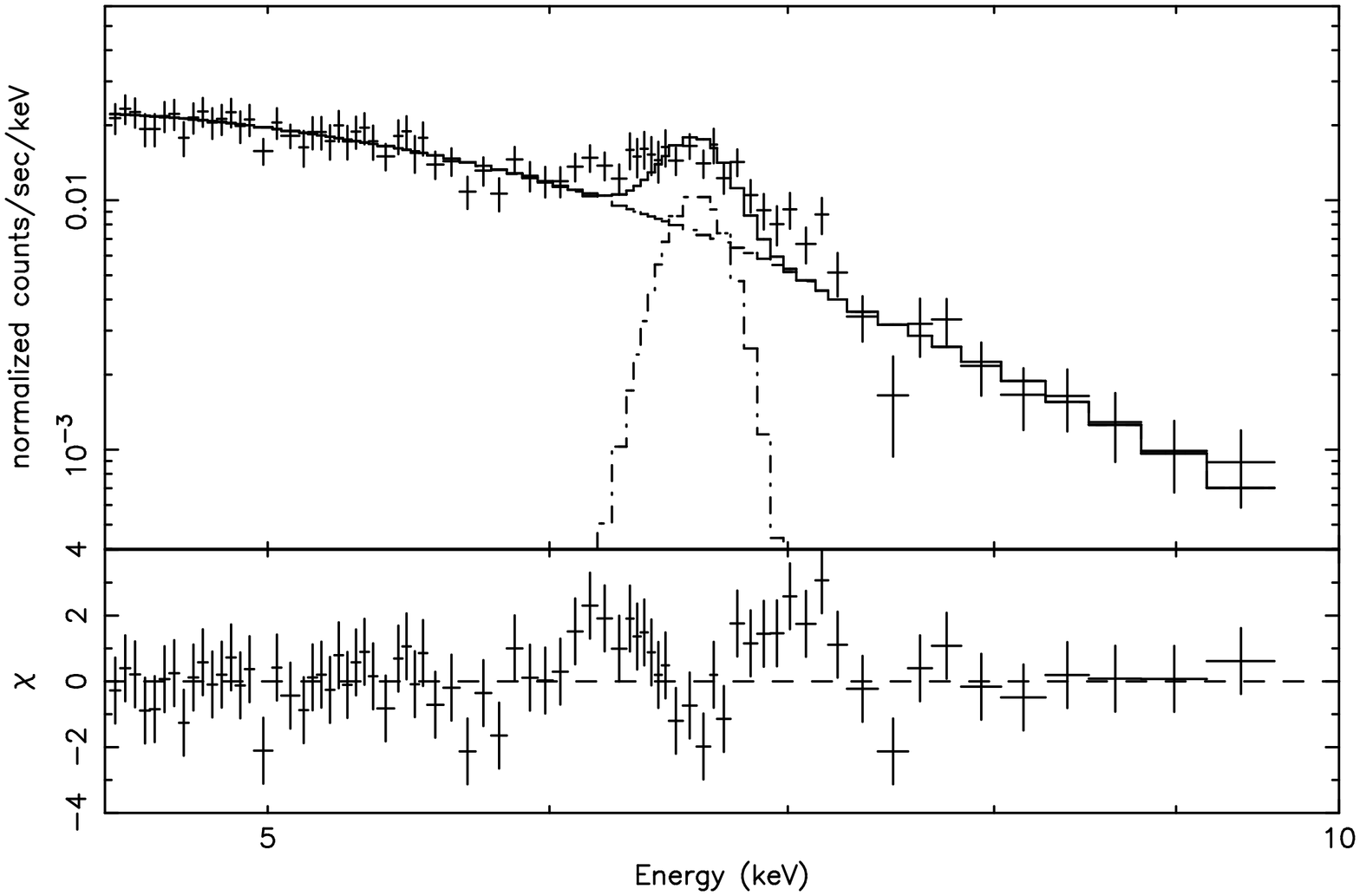}
\end{minipage}
\begin{minipage}{4.4cm}
{\bf b. single broad}\\
\vspace*{-1.5cm}
\includegraphics[height=5.8cm,width=4.4cm,angle=-90]{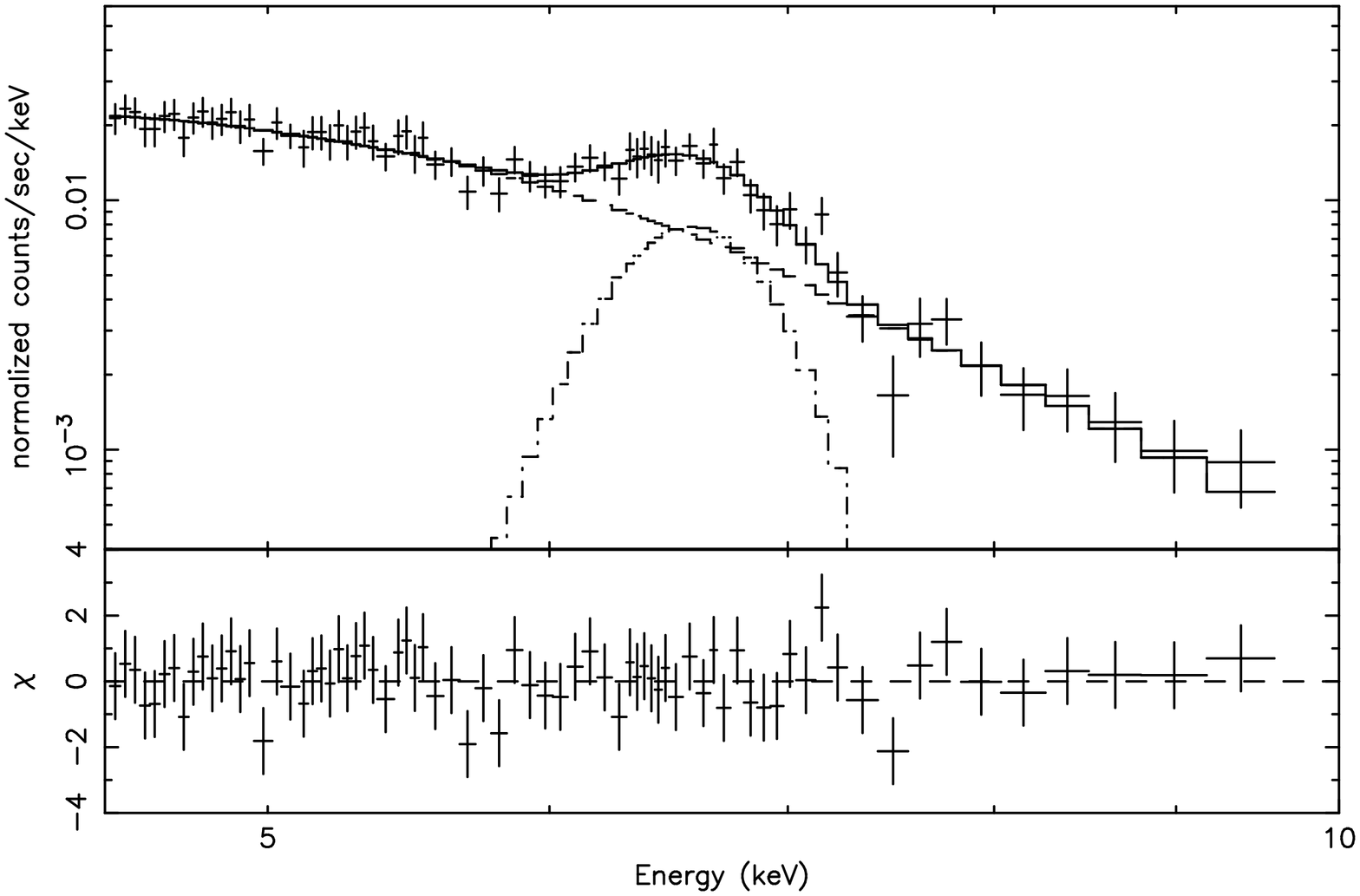}
\end{minipage}
\begin{minipage}{4.4cm}
{\bf c. double narrow}\\
\vspace*{-1.5cm}
\includegraphics[height=5.8cm,width=4.4cm,angle=-90]{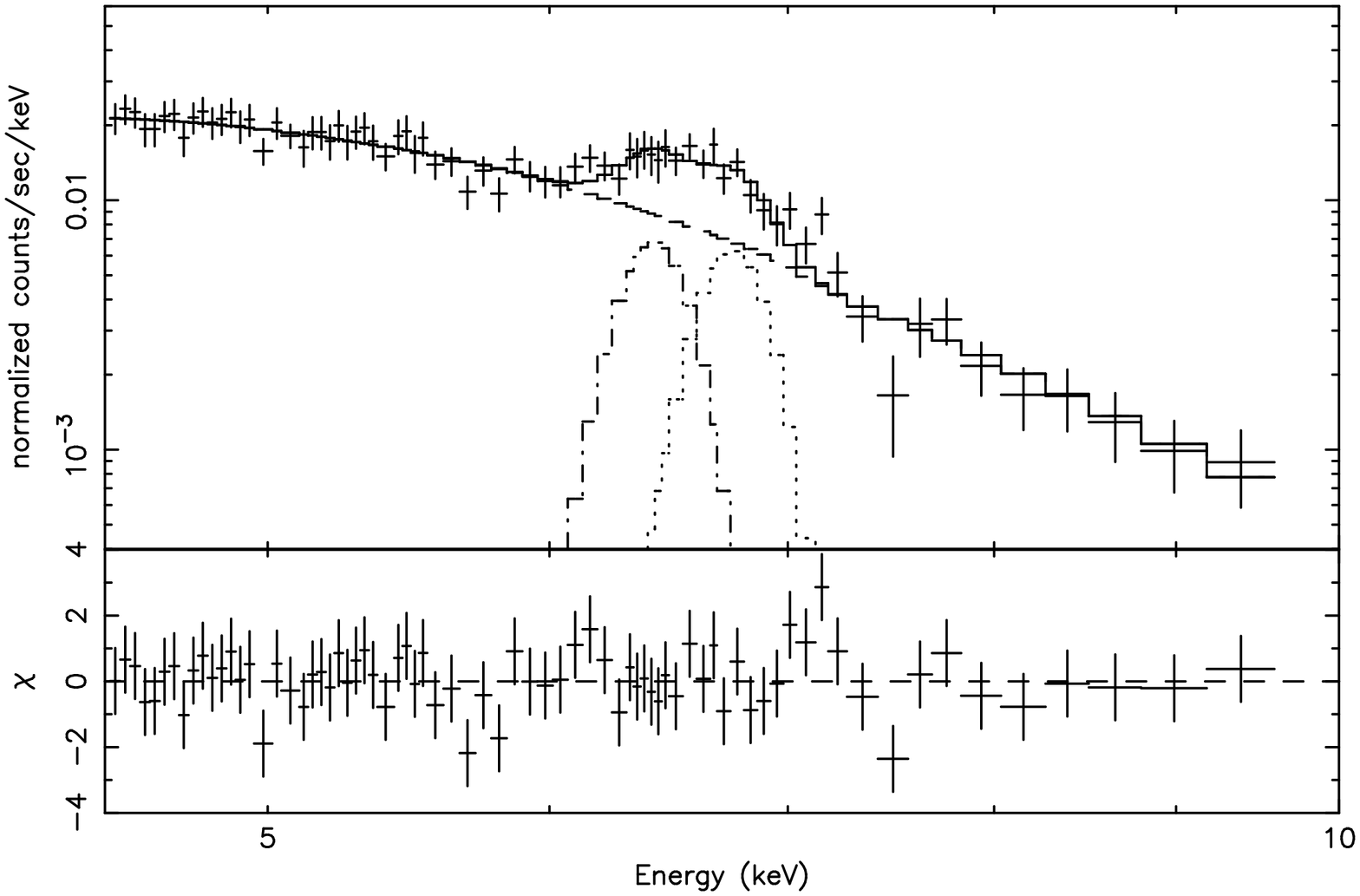}
\end{minipage}
\caption{Phase averaged SIS spectra in the iron K-line energy band,
fitted with a single narrow Gaussian (left), a single broad Gaussian (center),
and a double narrow Gaussian model (right).
The continuum spectrum is represented 
by a single temperature and a single column density
in the 4.5 -- 10.0 keV band (narrow band fitting).
The best-fit parameters are shown in table \ref{tbl:fit_line_first}.}
\label{fig:spec_line_single}
\end{figure*}

We attempted to quantify the 0.8 -- 10.0 keV continua,
neglecting for the moment the line energy band of 6.0--7.2 keV.
However, the simplest spectral model for polars, 
namely a single temperature bremsstrahlung continuum
absorbed by one single column density,
failed to reproduce either of the observed spectra
(figure \ref{fig:full_spec_discarded_fit} left;
Model 1 in table \ref{tbl:fit_continuum}).
This failure is not surprising, considering that
polars generally exhibit multi-temperature hard X-ray emission
with complex absorption by the pre-shock absorber (Norton and Watson 1989).
The extremely high temperature
obtained by this simple fitting, $>200$ keV, is presumably an artifact,
compared with the temperature of $14.7$ keV measured with Ginga.

To accurately estimate the hottest component of the continuum, 
avoiding complex absorption in soft energies,
we then restricted the fit energy band to a
narrower hard energy band of 4.5 -- 10.0 keV
(Model 2 in table \ref{tbl:fit_continuum}).
This lower limit (4.5 keV) was determined
in a way described in $\S 3.2$ by Ezuka and Ishida (1994).
In this case, $N_{\mbox{\tiny H}}$ is determined solely by the depth
of the iron K-edge absorption at $\sim$ 7.1 keV.
This model has been successful on the spectra of both phases,
yielding a temperature consistent with the Ginga value.
Considering that 
the K-edge absorption in the observed spectra are relatively shallow,
the value of $N_{\mbox{\tiny H}}$ obtained in this way
is thought to approximate the covering-fraction-weighted mean value
of multi-valued absorption.
The mean value of $N_{\mbox{\tiny H}}$ in the accretion column
is hence inferred to be $\sim 10^{23}$ cm$^{-2}$.

We next fitted the original 0.8 -- 10.0 keV spectra
(but excluding the iron K line region)
by a two-temperature bremsstrahlung with single $N_{\mbox{\tiny H}}$,
and obtained acceptable results
(right panel of figure \ref{fig:full_spec_discarded_fit} and
Model 3 in table \ref{tbl:fit_continuum}).
However, the first temperature $kT_{1}$ is still poorly determined;
the fit became unacceptable for the pole-on spectra
when we fixed $kT_{1}$ to the Ginga value.
Furthermore, the obtained $N_{\mbox{\tiny H}}$ is
about $1 \times 10^{22}$ cm$^{-2}$ in either case,
which is not consistent with the inference from the narrow band fitting.
Thus, we regard Model 3 as inappropriate.

A fourth spectral model we employed
consists of a single temperature bremsstrahlung
and double-valued photoelectric absorption
($N_{\mbox{\tiny H1}}$ and $N_{\tiny H2}$),
which is so-called partial covered absorption model (PCA model).
This model has been fully acceptable for both phases
(Model 4 in table \ref{tbl:fit_continuum}).
The obtained $N_{\mbox{\tiny H2}}$ is consistent with
that suggested by the narrow band fitting, and
the fit remained good even when we fix the temperature to the Ginga value.
We therefore utilize this model (Model 4) as the best representation
of the continua for both phases. 
The solid curves in figure \ref{fig:full_band_spec_best}
refer to this modeling.

\begin{table*}
\caption{Best Fit continuum parameters for the GIS and SIS spectra of V834 Cen$^{a}$.}

\begin{center}
\begin{tabular}{lcccccc}
\hline 
{Model$^{b}$}&	{$kT_{1}$}&	{$kT_{2}$}&	{$N_{\mbox{\tiny H1}}$}&	{$N_{\mbox{\tiny H2}}$}&	{Cov.\ Frac$^{c}$}&{$\chi^2_{\nu}$ (dof)}\\
{}&		{ keV }&	{ keV }&	{$\times 10^{22}$ cm$^{-2}$}&	{$\times 10^{22}$ cm$^{-2}$}&	{ \% }&		{}\\
\hline \hline 
\multicolumn{7}{l}{\bf POLE\_ON phase}\\
{Model 1}&	{$>200$}	&{--}&		{$0.12^{+0.02}_{-0.02}$}&	{--}&				{--}&		{2.19 (188)}\\
{Model 2$^{d}$}&{$10.8^{+26.8}_{-4.8}$}	&{--}&	{$14.5^{+8.1}_{-8.0}$}&		{--}&				{--}&		{0.49 (94)}\\
{Model 3}&	{$>158$}	&{$0.19^{+0.01}_{-0.01}$}&			{$0.82^{+0.08}_{-0.08}$}&{--}&	{--}&		{1.01 (186)}\\
{}&		{$14.7^{e}$}	&{$0.15^{+0.01}_{-0.01}$}&			{$1.21^{+0.08}_{-0.08}$}&{--}&	{--}&		{1.64 (187)}\\
{Model 4}&	{$>164$}	&{--}&		{$<0.02$}&			{$7.41^{+2.26}_{-1.70}$}&{$41.3^{+3.4}_{-3.3}$}&{0.91 (186)}\\
{}&		{$14.7^{e}$}	&{--}&		{$<0.08$}&			{$8.83^{+1.90}_{-1.59}$}&{$60.4^{+2.4}_{-2.5}$}&{1.00 (187)}\\
\multicolumn{7}{l}{\bf SIDE\_ON phase}\\
{Model 1}&	{$>171$}&	{--}&		{$<0.04$}&			{--}&				{--}&		{1.00 (174)}\\
{Model 2$^{d}$}&{$12.4^{+180}_{-7.1}$}&{--}&	{$13.9^{+11.3}_{-11.2}$}&	{--}&				{--}&		{0.47 (70)}\\
{Model 3}&	{$>200$}&	{$0.25^{+0.03}_{-0.03}$}&{$0.36^{+0.12}_{-0.13}$}&{--}&				{--}&		{0.85 (172)}\\
{}&		{$14.7^{e}$}&	{$0.18^{+0.01}_{-0.01}$}&{$0.78^{+0.10}_{-0.10}$}&{--}&				{--}&		{1.135 (173)}\\
{Model 4}&	{$>49.3$}&	{--}&		{$<0.02$}&			{$12.7^{+12.5}_{-1.2}$}&{$35.6^{+10.0}_{-14.1}$}&{0.81 (172)}\\
{}&		{$14.7^{e}$}&{--}&		{$<0.06$}&			{$11.2^{+4.5}_{-3.4}$}&	{$50.7^{+5.4}_{-5.1}$}&	{0.85 (173)}\\
\multicolumn{7}{l}{\bf Phase average}\\
{Model 2$^{d}$}&{$11.7^{+15.2}_{-4.4}$}&{--}&	{$14.0^{+5.8}_{-4.2}$}&		{--}&				{--}&		{0.49 (122)}\\
{Model 4}&	{$14.7^{e}$}&{--}&		{$<0.06$}&			{$9.79^{+1.69}_{-1.47}$}&	{$56.9^{+2.1}_{-2.1}$}&{1.07 (275)}\\
\hline 
\end{tabular}

\begin{tabular}{cl}
{$^{a}$}&{Excluding the Fe K$_{\alpha}$ line band (6.0 -- 7.2 keV).}\\
{$^{b}$}&{Model 1/2: single $N_{\mbox{\tiny H}}$, single $kT$. Model 3  : single $N_{\mbox{\tiny H}}$, multi $kT$.}\\
{}	&{Model 4: multi  $N_{\mbox{\tiny H}}$, single $kT$.}\\
{$^{c}$}&{The covering fraction (\%) of $N_{\mbox{\tiny H1}}$.}\\
{$^{d}$}&{Fitting in the 4.5 -- 10.0 keV band. For other models, the 0.8 -- 10.0 keV band is used.}\\
{$^{e}$}&{Continumn temparature fixed at the value measured with Ginga (Ishida et al. 1991).}\\
\end{tabular}
\end{center}
\end{table*}


\subsubsection{ The Fe-K lines }


Having quantified the continuum spectra of V834 Cen,
we proceed to the study of the iron K-line.
For this purpose, we employ the phase-averaged spectrum,
and again limit the energy range to the 4.5 -- 10.0 keV band
to avoid complex absorption structure in lower energies.
As a first-cut attempt, 
we modeled the iron line with a single Gaussian model,
while represented the continuum with a single-temperature bremsstrahlung
absorbed with a single column density (to reproduce the iron K edge),
but the fit failed to reproduce the line profile as shown
in figure \ref{fig:spec_line_single}a.
A broad Gaussian model with $\sigma \sim 0.28$ keV
has been found to be successful
(table \ref{tbl:fit_line_first}; figure \ref{fig:spec_line_single}b),
but the obtained line centroid energy is too low for ionized Fe-K species
that are expected for a plasma of temperature $\sim$ 10 keV.
We can alternatively fit the data successfully with two narrow Gaussians
(table \ref{tbl:fit_line_first}; figure \ref{fig:spec_line_single}c),
where the centroid energy of the first Gaussian turned out to be consistent
with that of the fluorescent iron K$_{\alpha}$ line (6.40 keV);
that of the second Gaussian comes in between those of
He-like iron K$_{\alpha}$ line (6.65 -- 6.70 keV) and H-like line (6.97 keV).
This means that the second Gaussian is in reality a composite of
the H-like and He-like lines.

\begin{figure}
\centerline{\includegraphics[width=8cm]{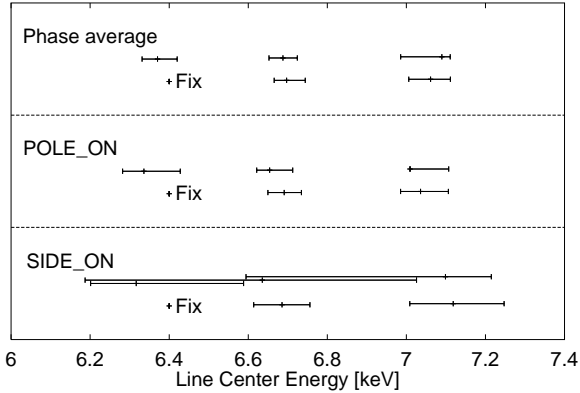}}
\caption{The best-fit line centroid energies with 68 \% errors
in terms of the three narrow Gaussian model.
From the top to bottom panels, the results for the phase average,
pole-on phase, and side-on phase are shown.
The upper plot in each panel shows the results
when three centroid energies are left free.
The lower plot shows the results when the centroid energy of
the first Gaussian is fixed at 6.40 keV.
The fit energy range is 4.5 -- 10.0 keV.}
\label{fig:spec_line_center}
\end{figure}

We therefore employed a line model consisting of three narrow Gaussian,
each having a free centroid energy and a free normalization.
We have then obtained an acceptable fit,
with the three centroid energies consistent with the Fe-K line energies
of the neutral, He-like, and hydrogen-like spices,
as shown in figure \ref{fig:spec_line_center}.
These three lines have been observed in the spectra
of many polars with ASCA (Ezuka and Ishida 1999).
Hereafter, we adopt the three narrow Gaussian model
in quantifying the iron lines of V834 Cen.

\begin{table*}
\caption{Best Fit parameters of the iron K$_{\alpha}$ line$^{a}$.}
\begin{center}
\begin{tabular}{lccccc}
\hline
{Line model}&	\multicolumn{4}{c}{iron K$_{\alpha}$ line}&	{statistics}\\
\hline
{}&		{L.C. 1$^{b}$}&	{$\sigma_{\mbox{\tiny 1}}$}&	{L.C. 2$^{b}$}&	{$\sigma_{\mbox{\tiny 2}}$}& {$\chi^2_{\nu}$ (dof)}\\
{}&		{(keV)}			&{(keV)}		&{(keV)}		&{(keV)}	&{}\\
\hline \hline
{single narrow}&{$6.63^{+0.04}_{-0.05}$}&{$0.0001^{c}$}&		{--}&			{--}&		{1.37 (69)}\\
{single broad} &{$6.66^{+0.05}_{-0.05}$}&{$0.29^{+0.06}_{-0.05}$}&	{--}&			{--}&		{0.66 (68)}\\
{double narrow}&{$6.45^{+0.08}_{-0.11}$}&{$0.0001^{c}$}&		{$6.79^{+0.23}_{-0.46}$}&{$0.0001^{c}$}&{0.91 (67)}\\
\hline
\end{tabular}
\begin{tabular}{cl}
{$^{a}$}&{The phase averaged spectrum. Only SIS data are used for the fitting.}\\
{}	&{Fitted with a single temparature and single column density model in 4.5--10.0 keV.}\\
{}	&{Fixed the continuum temperature and $N_{\mbox{\tiny H}}$ to the Model2 in table \ref{tbl:fit_continuum}}\\
{$^{b}$}&{Line center energies.}\\
{$^{c}$}&{Fixed.}\\
\end{tabular}
\end{center}
\label{tbl:fit_line_first}
\end{table*}

\begin{figure*}
\begin{minipage}{6.7cm}
{\bf Pole-on spectrum}\\ 
\vspace*{-1.6cm}

\centerline{\includegraphics[height=8cm,angle=-90]{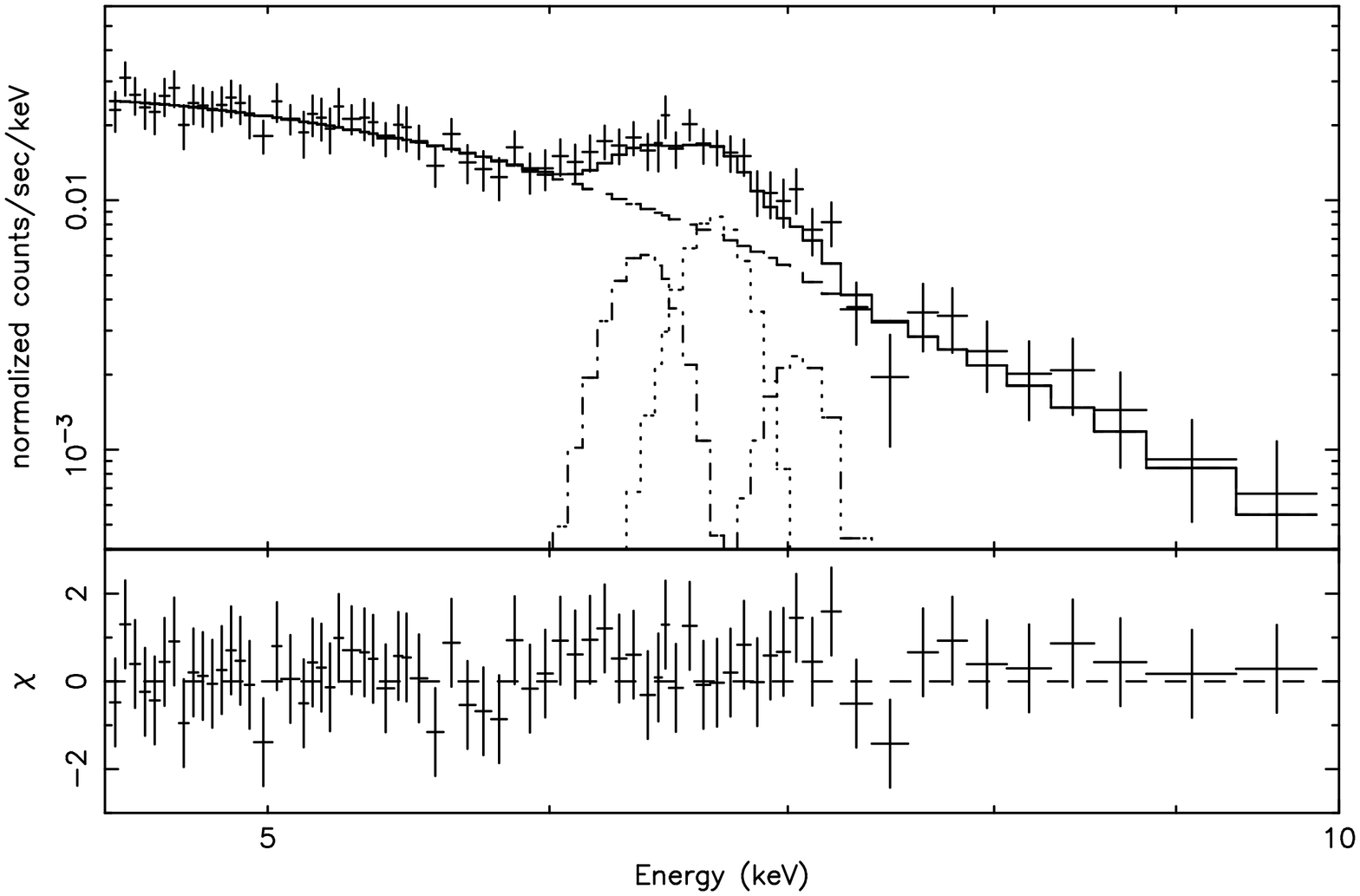}}
\end{minipage}
\begin{minipage}{6.7cm}
{\bf Side-on spectrum}\\
\vspace*{-1.6cm}

\centerline{\includegraphics[height=8cm,angle=-90]{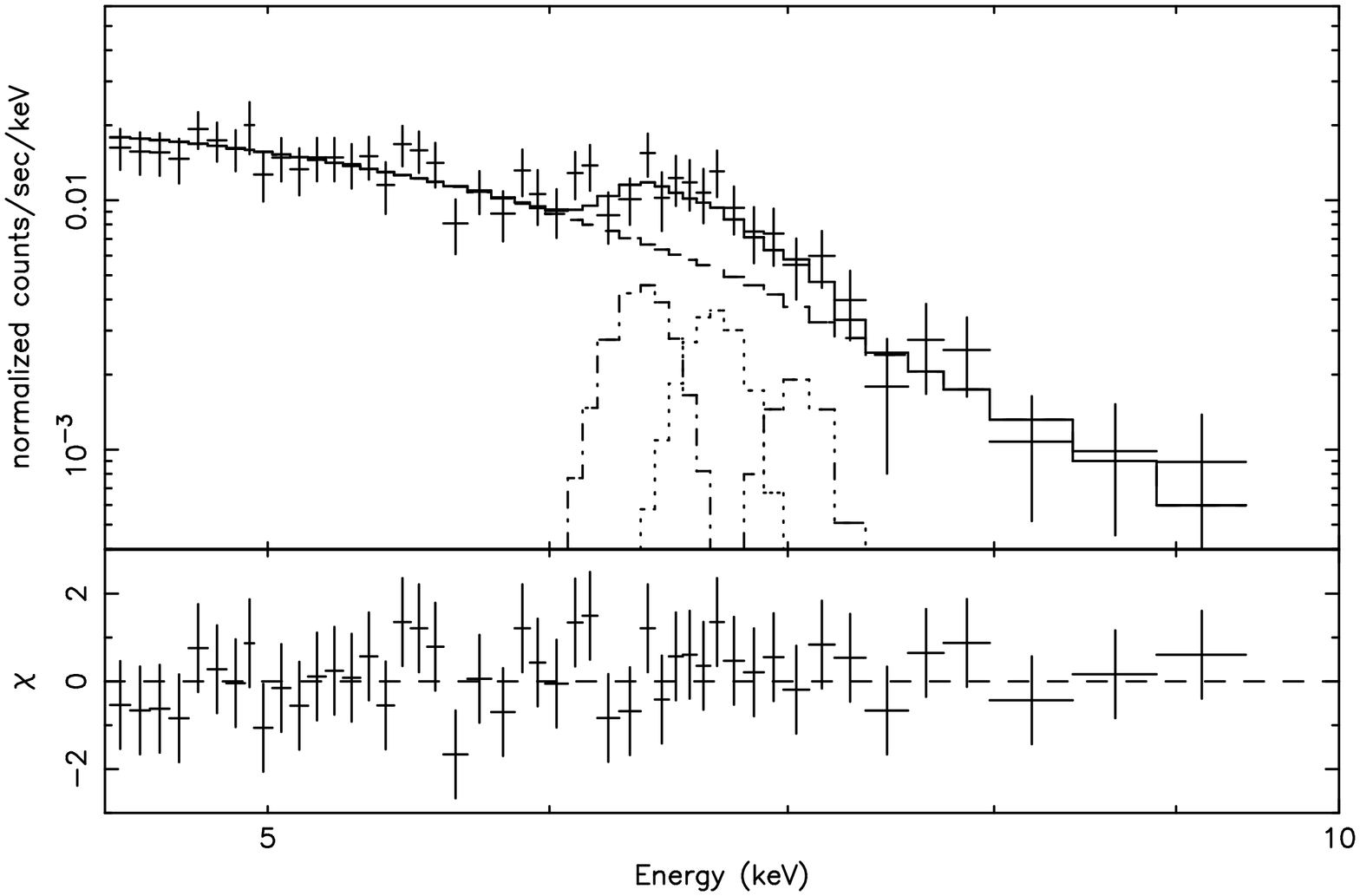}}
\end{minipage}
\caption{Phase resolved spectra over the Fe-K line energies,
fitted with three narrow Gaussians.
The central energy of the first Gaussian is fixed at 6.4keV.
The best-fit parameters are shown in table \ref{tbl:fit_line}.}
\label{fig:spec_line_phase}
\end{figure*}

As our final analysis,
we have repeated the three-Gaussian fitting to the spectra,
by fixing the line centroid energy of the first Gaussian
(identified to a fluorescent line) at 6.40 keV.
The result is of course successful,
and the obtained parameters are given in table \ref{tbl:fit_line}
as well as figure \ref{fig:spec_line_center}.
Figure \ref{fig:spec_line_phase} shows the phase-resolved SIS spectra
fitted with this model.
For consistency check,
we have expanded the fit energy range back to 0.7 -- 10.0 keV,
and performed full-band fitting,
employing the continuum Model 4
and the three narrow Gaussian for the iron K-line.
The results, presented in table \ref{tbl:fit_line},
are generally consistent with those from the narrow-band analysis.

The temperature and $N_{\mbox{\tiny H}}$
obtained in these final fits are thus the same between the two phases
within errors,
and the EW of fluorescent and H-like lines are also consistent
with being unmodulated.
In contrast, the EW of the He-like line is enhanced by
$(\xi_{\mbox{\tiny pole}}/\xi_{\mbox{\tiny side}})^{\mbox{\tiny obs}}
\equiv 1.87_{-0.54}^{+0.54}$ times
(narrow band fitting) or $1.91_{-0.64}^{+0.47}$ times (full band fitting)
in the pole-on phase compared to the side-on phase,
with $>90$ \% statistically significance.

\begin{table*}
\caption{Best Fit parameters of the iron K$_{\alpha}$ line with three narrow gaussans.}
\begin{center}
\begin{tabular}{lcccccccc}
\hline 
{Phase$^{a}$}&	\multicolumn{2}{c}{continuum}&\multicolumn{1}{c}{Fluorescent}&	\multicolumn{2}{c}{He-like}&	\multicolumn{2}{c}{H-like}  &{$\chi^2_{\nu}$ (dof)}\\
{}&		{kT}&{N$_{\mbox{\tiny H}}$}	&{EW}&	{l.\ c.\ $^{b}$}&	{EW}&	{l.\ c.\ $^{b}$}&	{EW}&{}\\
{}&		{(keV)}&{$\times 10^{22}$ cm$^{-2}$}&	{(eV)}&	{(keV)}&{(eV)}&			{(keV)}&{(eV)}&		{}\\
\hline \hline 
\multicolumn{8}{l}{\bf Narrow Band Fitting$^{c}$}\\
{Average}&{$9.3^{+10.4}_{-2.9}$}&{$15.4^{+5.9}_{-5.8}$}& 	{$244^{+59}_{-83}$}&	{$6.70^{+0.06}_{-0.04}$}& 	{$352^{+90}_{-91}$}	&{$7.07^{+0.08}_{-0.44}$}& {$192^{+81}_{-97}$} &{0.67 (154)}\\
{Pole-on}&{$9.6^{+20.3}_{-2.87}$}&{$15.6^{+8.0}_{-6.5}$}&	{$242^{+87}_{-102}$}&	{$6.69^{+0.07}_{-0.05}$}&	{$417^{+104}_{-113}$}	&{$7.04^{+0.11}_{-0.15}$}& {$223^{+123}_{-118}$}&{0.59 (140)}\\
{Side-on}&{$9.0^{+50.0}_{-1.0}$}&{$16.6^{+13.4}_{-9.5}$}&	{$270^{+122}_{-175}$}&	{$6.69^{+0.70}_{-0.13}$}&	{$223^{+94}_{-101}$}	&{$7.14^{+0.16}_{-0.48}$}& {$223^{+123}_{-89}$}&{0.64 (103)}\\
\multicolumn{8}{l}{\bf Full Band Fitting$^{d}$}\\
{Average}&\multicolumn{2}{c}{}&	{$347^{+53}_{-52}$}&	{$6.71^{+0.05}_{-0.06}$}&{$456^{+57}_{-70}$}&{$7.07^{+0.10}_{-0.08}$}&{$230^{+88}_{-56}$}&{1.05 (327)}\\
{Pole-on}&\multicolumn{2}{c}{(Model 4 in table \ref{tbl:fit_continuum})$^{e}$}&		{$229^{+164}_{-17}$}&	{$6.71^{+0.07}_{-0.10}$}& {$650^{+19}_{-164}$}	&{$7.06^{+0.12}_{-0.15}$}& {$282^{+58}_{-164}$}&{0.98 (243)}\\
{Side-on}&\multicolumn{2}{c}{}&	{$368^{+76}_{-119}$}&	{$6.71^{+0.62}_{-0.13}$}&	{$341^{+115}_{-111}$}	&{$7.13^{+0.21}_{-0.56}$}& {$222^{+199}_{-105}$}&{0.82 (210)}\\
\hline 
\end{tabular}

\medskip

\begin{tabular}{cl}
{$^{a}$}&{Pole-on phase: $\phi = 0.73 \pm 0.25$. Side-on phase: $\phi = 0.23 \pm 0.25$.}\\
{$^{b}$}&{The line center (keV). That of the fluorescent component is fixed at 6.40 keV.}\\
{$^{c}$}&{The determination of continumn spectrum is performed in 4.5 keV -- 10.0 keV}\\
{}	&{with a single $N_{\mbox{\tiny H}}$ and single temparature model.}\\
{$^{d}$}&{The determination of continumn spectrum is performed in 0.8 keV -- 10.0 keV}\\
{}	&{with a multi $N_{\mbox{\tiny H}}$ and single temparature model (Model 4 in table \ref{tbl:fit_continuum}).}\\
{$^{e}$}&{The temparature is fixed to the Ginga value. Continuum parameters are fixed.}\\
\end{tabular}
\end{center}
\label{tbl:fit_line}
\end{table*}


\section{Discussion}
\label{section:discussion}

\subsection{Origin of the line intensity modulation in V834 Cen}
\label{section:origin}

In order to verify our interpretation,
mentioned in section \ref{section:mechanism},
we observed the polar V834 Cen with ASCA.
The EW of the He-like iron-K$_\alpha$ line has been confirmed
to be enhanced by a factor of $(\xi_{\mbox{\tiny pole}}/\xi_{\mbox{\tiny side}})^{\mbox{\tiny obs}} = 1.87 \pm 0.54$
in the pole-on phase relative to the side-on phase.
Can we explain the observed rotational modulation of the He-like iron line EW
of V834 Cen by some conventional mechanisms?
An obvious possibility is
that the temperature variation modulates the iron line intensity.
However, the continuum temperature determined by the narrow band fitting
(table \ref{tbl:fit_line}) is almost constant within the error
to explain the rotational modulation of the line intensity ratio.
Alternatively, the bottom part of the accretion column,
where the emissivity of the He-like iron line is highest
(figure \ref{fig:column_emis}),
may be eclipsed by the WD surface
so that the He-like line is reduced in the side-on phase.
However, such a condition does not occur
under the geometrical parameters of V834 Cen.
Furthermore, we do not see any dip
in the rotation-folded hard X-ray light curve.
Therefore, this explanation is not likely, either.
A third possibility is that a part of iron line photons come from
some regions other than the accretion column.
In fact, X-rays from polars are known to be
contributed by photons reflected from the WD surface
(Beardmore et al. 1995; Done \& Beardmore 1995; Done \& Magdziarz 1998).
This mechanism can explain the production of fluorescent lines,
and possibly its rotational modulation,
but cannot produce highly ionized iron lines.
From these considerations, we conclude
that the observed iron line modulation is difficult to account for
without appealing to the resonance scattering effects.

\begin{figure}
\includegraphics[width=7cm]{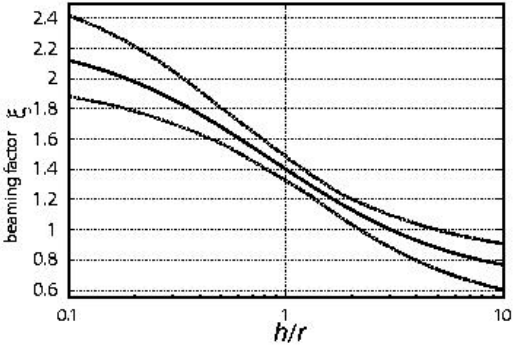}
\caption{
The expected line beaming factor
$\xi_{\mbox{\tiny pole}}/\xi_{\mbox{\tiny side}}$
of V834 Cen due to pure geometrical effects
at the limit of large optical depth,
calculated by considering the system geometry of V834 Cen
and the exposure for each phase.
It is shown as a function of $h/r$.
The solid line corresponds to the most likely system geometry of
$(i, \beta)=(45^\circ, 25^\circ)$,
while dashed lines reflect uncertainties in the system geometry.}
\label{fig:geometrical_beaming_factor}
\end{figure}

Our next task is to examine
whether the geometrical collimation mechanism
(\S~\ref{section:mechanism_geometrical_beaming}) can explain the observation.
When the optical depth of resonance scattering is very high
and hence
the line photons come solely from the surface of the accretion column,
the angular distribution of the iron line intensity is given analytically by
equation (\ref{eq:geometrical_profile})
in Appendix \ref{section:geometrical_beaming}.
By averaging this distribution under the geometry of V834 Cen,
and taking into account the exposure for the two phases,
we have calculated the expected line beaming factor $\xi$ for the ASCA data
as shown in Figure \ref{fig:geometrical_beaming_factor}.
Thus, to explain the observed enhancement of He-like line
solely by the geometrical beaming,
a very flat coin-shaped column with $h/r \sim 0.2$ would be required.
Furthermore, the observed $\xi$ should be lower by about 30--40\%
than the ideal calculation in Figure \ref{fig:geometrical_beaming_factor},
because finite optical depths,
expected under a typical plasma density of polar accretion column,
reduce the enhancement (Figure \ref{fig:sim_thin_plasma}).
We hence conclude that the geometrical mechanism alone is insufficient
to explain the observed He-like iron line enhancement in V834 Cen
(\S~\ref{section:mechanism}),
and hence the additional collimation due to the velocity gradient effect is
needed.

Then, are all the observed results consistently explained in our picture
that incorporates the geometrical and velocity-gradient effects?
At the temperature of V834 Cen ($\sim$ 15 keV),
the resonant photons in fact contribute about 65\%
to the observed He-like iron K$_\alpha$ line,
the rest coming from intercombination  ($\sim 20$\%)
and forbidden  ($\sim 15$\%) lines
which are free from the resonance scattering effects
(figure \ref{fig:line_pow1}).
As a result, the line collimation is expected to be
somewhat weakened in the ASCA spectra,
where we cannot separate these unmodulated lines from the resonance line.
After correcting for this reduction,
the true value of $\xi$ for the He-like resonance line
is calculated to be $\sim 2.3$.
This is within the range that can be explained by our scenario.

How about the H-like iron K$_\alpha$ line?
Even though it consists entirely of resonant photons,
its modulation in the V834 Cen data
has been insignificant (table \ref{tbl:fit_line}).
Presumably,
this is mainly due to technical difficulties in detecting this weak line,
under the presence of the stronger He-like line adjacent to it.
Furthermore, we expect the hydrogen-like iron line to be
intrinsically less collimated than the He-like resonance iron line,
because the H-like line photons are produced predominantly
in the top regions of the accretion column (figure \ref{fig:column_emis}):
there, the electron density is lower, thermal Doppler effect is stronger,
and the path of escape from the column is shorter,
as compared to the bottom region where the He-like lines are mostly produced.
Therefore, the H-like line is expected to be less collimated
than the He-like resonance line (figure \ref{fig:sim_v834Cen_beaming}).
Note in the observation
blended He-like line is almost the same collimation as H-like line.

From these considerations, we conclude
that the ASCA results on V834 Cen
can be interpreted consistently by our POLE scenario.

\subsection{Determination of the accretion column parameters}
\label{section:determine}
The line beaming effect we have discovered is expected to provide
unique diagnostics of the accretion column of magnetic WDs.
The physical condition in the accretion column
is described by four parameters:
$kT^{\mbox{\tiny sh}}$, $h$, $r$, and $n_e^{\mbox{\tiny sh}}$.
To determine these parameters,
four observational or theoretical constraints are required.
Usually, observations provide two independent quantities,
the temperature $kT_{\mbox{\tiny OBS}}$
and the volume emission measure VEM, which is 
$\sim (n_e^{\mbox{\tiny sh}})^2 \cdot h r^2$.
Also there is one theoretical constraint,
that the shock heated plasma cools only by the free-free cooling,
which relates $kT^{\mbox{\tiny sh}}$, $n_e^{\mbox{\tiny sh}}$,
and $h$ as in equation (\ref{eq:column_height}).

With these three constraints,
and taking into account the Aizu's solution (1973),
we can express $h$ and $r$ as
\begin{eqnarray}
h &=& 1.61 \times 10^7 \mbox{cm}
\left(\frac{kT^{\mbox{\tiny sh}}}{17.7 \mbox{keV}}\right)^{\frac{1}{2}}
\left(\frac{n_e^{\mbox{\tiny sh}}}{10^{16} \mbox{cm$^{-3}$}}\right)^{-1}
\label{eq:v834cen_h}\\
r &=& 2.15 \times 10^7 \mbox{cm}
\left(\frac{\mbox{VEM}}{1.30\times 10^{54} \mbox{cm$^{-3}$}}\right)
\left(\frac{kT^{\mbox{\tiny sh}}}{17.7 \mbox{keV}}\right)^{-\frac{1}{4}}
\nonumber \\ && \times 
\left(\frac{n_e^{\mbox{\tiny sh}}}{10^{16} \mbox{cm$^{-3}$}}
\right)^{-\frac{1}{2}}.
\label{eq:v834cen_r}
\end{eqnarray}
Here, we normalized the VEM to the value of $1.30\times 10^{54}$ cm$^{-3}$
obtained from V834 Cen, adopting a distance of 86 pc (Warner 1987).
The value of $kT^{\mbox{\tiny sh}} = 17.7$ keV was determined
from the expected mass of WD
by the observed ratio of H-like to He-like iron K$_\alpha$ lines,
considering the vertical temperature gradient (Ezuka and Ishida 1999 figure 5);
it is consistent with an independent calculation by Wu et al (1995)
based on the Ginga observation (Ishida 1991).
When $n_e^{\mbox{\tiny sh}}$ is low,
the solutions to $h$ and $r$ imply a long cylinder-like column,
while a flat coin-shaped geometry is indicated by
high values of $n_e^{\mbox{\tiny sh}}$.
However, due to the lack of one more piece of information,
the value of $n_e^{\mbox{\tiny sh}}$ has so far been left undetermined.
As a consequence, we have not been able to determine the column geometry.

\begin{figure}
\includegraphics[height=5.5cm]{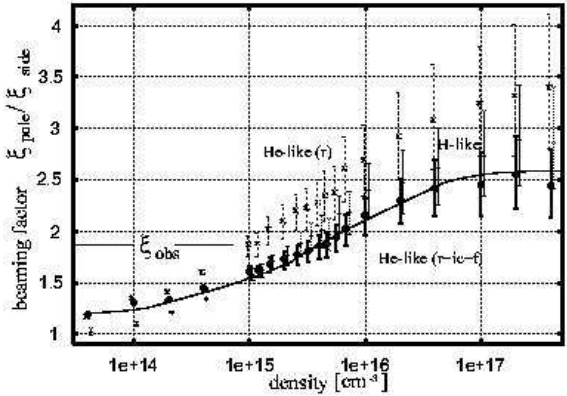}
\caption{The expected beaming factor
$\xi_{\mbox{\tiny pole}}/\xi_{\mbox{\tiny side}}$
of V834 Cen, calculated by Monte-Carlo simulations
under the actual observing condition with ASCA,
and shown as a function of the assumed post-shock density.
The shock temperature is set to 17.7 keV,
and the volume emission measure is set to $1.30 \times 10^{54}$ cm $^{-3}$.
The dashed and dotted data points indicate $\xi$ of
He-like and H-like resonance iron K$_\alpha$ photons, respectively.
The solid data points show $\xi$ of the blended  He-like iron line,
which includes intercombination and forbidden lines.
The error bars take into account those of the geometrical parameters.}
\label{fig:sim_v834Cen_beaming}
\end{figure}

The present work provides us with the needed fourth information,
the value of $\xi$, which reflects the accretion column condition.
Suppose we specify a value of $n_e^{\mbox{\tiny sh}}$.
Then, equations (\ref{eq:v834cen_h}) and (\ref{eq:v834cen_r})
determine $h$ and $r$ respectively,
which in turn are used as inputs
to the Monte-Carlo simulation to predict $\xi$.
In this way, we have calculated $\xi$ for V834 Cen
as a function of $n_e^{\mbox{\tiny sh}}$,
and show the results in Figure \ref{fig:sim_v834Cen_beaming}.
By comparing it with the observed value of
$(\xi_{\mbox{\tiny pole}}/\xi_{\mbox{\tiny side}})^{\mbox{\tiny obs}}
\sim 1.87$,
we obtain the best estimate as
$n_e^{\mbox{\tiny sh}} \sim 4.6 \times 10^{15}$ cm$^{-3}$.
This in turn fixes the column height as $h \sim 3.5 \times 10^{7}$ cm
and the radius as $r \sim 3.2\times 10^{7}$ cm.
These values are considered typical for polars.

Currently, the errors on $\xi_{\mbox{\tiny pole}}/\xi_{\mbox{\tiny side}}$ are so large
that the value of $n_e^{\mbox{\tiny sh}}$ is uncertain
almost by an order of magnitude,
with the allowed range being
$1 \times 10^{14}$ -- $1 \times 10^{16}$ cm$^{-3}$.
However, our new method will provide a powerful tool
for next generation instruments
with a larger effective area and an improved energy resolution.

\begin{table*}
\caption{Geometrical parameters of randomly sampled polars, and values of $\zeta$ expected for them}
\begin{center}
\begin{tabular}{ccccl}
\hline
{object name}&	{$i^{\mbox{\tiny a}}$}&		{$\beta^{\mbox{\tiny b}}$}& {$\zeta^{\mbox{\tiny c}}$}&
\multicolumn{1}{c}{reference}\\
\hline
{BL Hyi}&	{$70\pm10$}&	{$153\pm10$}&	{0.70}&	{Cropper 1990}\\
{UZ For}&	{$\sim88$}&	{$\sim14$}&	{0.71}&	{Ferrario et al. 1989}\\
{VV Pup}&	{$76\pm6$}&	{$152\pm6$}&	{0.71}&	{Cropper 1990}\\
{ST LMi}&	{$64\pm5$}&	{$141\pm4$}&	{0.71}&	{Cropper 1990}\\
{AN UMa}&	{$65\pm20$}&	{$20\pm5$}&	{0.87}&	{Cropper 1990}\\
{QQ Vul}&	{$65\pm7$}&	{$\sim23$}&	{0.88}&	{Cropper 1990, Schwope et al. 2000, Catalan et al. 1999}\\
{V1309 Ori}&	{$\sim80$}&	{$\sim30$}&	{0.86}&	{Harrop-Allin et al. 1997}\\
{EF Eri}&	{$58\pm12$}&	{$27\pm18$}&	{0.96}&	{Cropper 1990}\\
{AR UMa}&	{$50\pm10$}&	{$23\pm13$}&	{1.05}&	{Szkody et al. 1999}\\
{WW Hor}&	{$\sim74$}&	{$\sim48$}&	{1.06}&	{Bailey et al. 1988}\\
{DP Leo}&	{$76\pm10$}&	{$103\pm5$}&	{1.06}&	{Cropper 1990}\\
{MR Ser}&	{$43\pm5$}&	{$38\pm5$}&	{1.09}&	{Cropper 1990}\\
{J1015+0904}&	{$55\pm5$}&	{$43\pm7$}&	{1.10}&	{Burwitz et al. 1998}\\
{V834 Cen}&	{$45\pm9$}&	{$25\pm5$}&	{1.11}&	{Cropper 1990}\\
{AM Her}&	{$52\pm5$}&	{$66\pm5$}&	{1.13}&	{Wickramasinghe et al. 1991, Ishida et al. 1997}\\
{EK UMa}&	{$56\pm19$}&	{$56\pm19$}&	{1.14}&	{Cropper 1990}\\
{VY For}&	{$9\pm 3$}&	{$\sim 9$}&	{1.66}&	{Beuermann et al. 1989}\\ 
\hline
\end{tabular}

\begin{tabular}{cl}
{$^{a}$}&	{Inclination angle (degrees).}\\
{$^{b}$}&	{Pole colatitude (degrees).}\\
{$^{c}$}&	{Expected beaming factor to the average flux, assuming}\\
{}&		{that the same collimation as the case of V834 Cen occurs}\\
{}&		{and that only single accretion column emitts.}\\
\end{tabular}
\end{center}
\label{tbl:polar_geometry_list}
\end{table*}

\subsection{Effects on the abundance estimates}
\label{section:abundance}
The iron abundance of other polars, measured by Ezuka \& Ishida (1999),
are subject to some changes
when we properly consider the resonance scattering effects.
For this purpose, we show in table \ref{tbl:polar_geometry_list} 
the geometrical parameters ($i$ and $\beta$) of 17 polars,
which have been randomly selected from currently-known $\sim$ 50 polars.
The expected enhancement \(\zeta \equiv \sum_{\mbox{\tiny phase}} \xi\)
are also listed with assumption that
the resonance lines are collimated as much as that in the case of V834 Cen.
Thus, the implied corrections to the iron abundances of these objects
are at most $\pm 30$ \%,
which is generally within the measurement error.
Polars with almost side-on or almost pole-on geometry, like POLEs,
are subject to a relatively large modification in the abundance estimates,
but such a geometry is a rare case (here, only VY For).
Therefore, the distribution of metal abundances of polars,
0.1 -- 0.8 solar, measured by Ezuka \& Ishida (1999)
is considered still valid.


\subsection{Mystery of POLEs}
\label{section:discuss_statistical}
Our Monte-Carlo simulation ($\S$\ref{section:monte_carlo}) and
the observed study of V834 Cen ($\S$\ref{section:observation})
consistently indicate that the resonance iron K lines are enhanced
by a factor of $\xi_{\mbox{\tiny m}} =$ 2 -- 2.5
in the axial direction of accretion column.
The effect is thought to be ubiquitous among polars,
because we have so far employed very typical conditions among them.
Consequently, if a polar with co-aligned magnetic axis ($\beta \sim 0$)
is viewed from rear the pole-on direction ($i \sim 0$),
we expect the iron K line EW to be persistently enhanced by 2 - 2.5 times.
When this enhancement is not considered,
such objects would yield artificially higher metallicity by similar factors.
We conclude that the three POLEs ($\S$\ref{section:intro_fe})
are exactly such objects.

\begin{figure*}
\includegraphics[height=6.5cm]{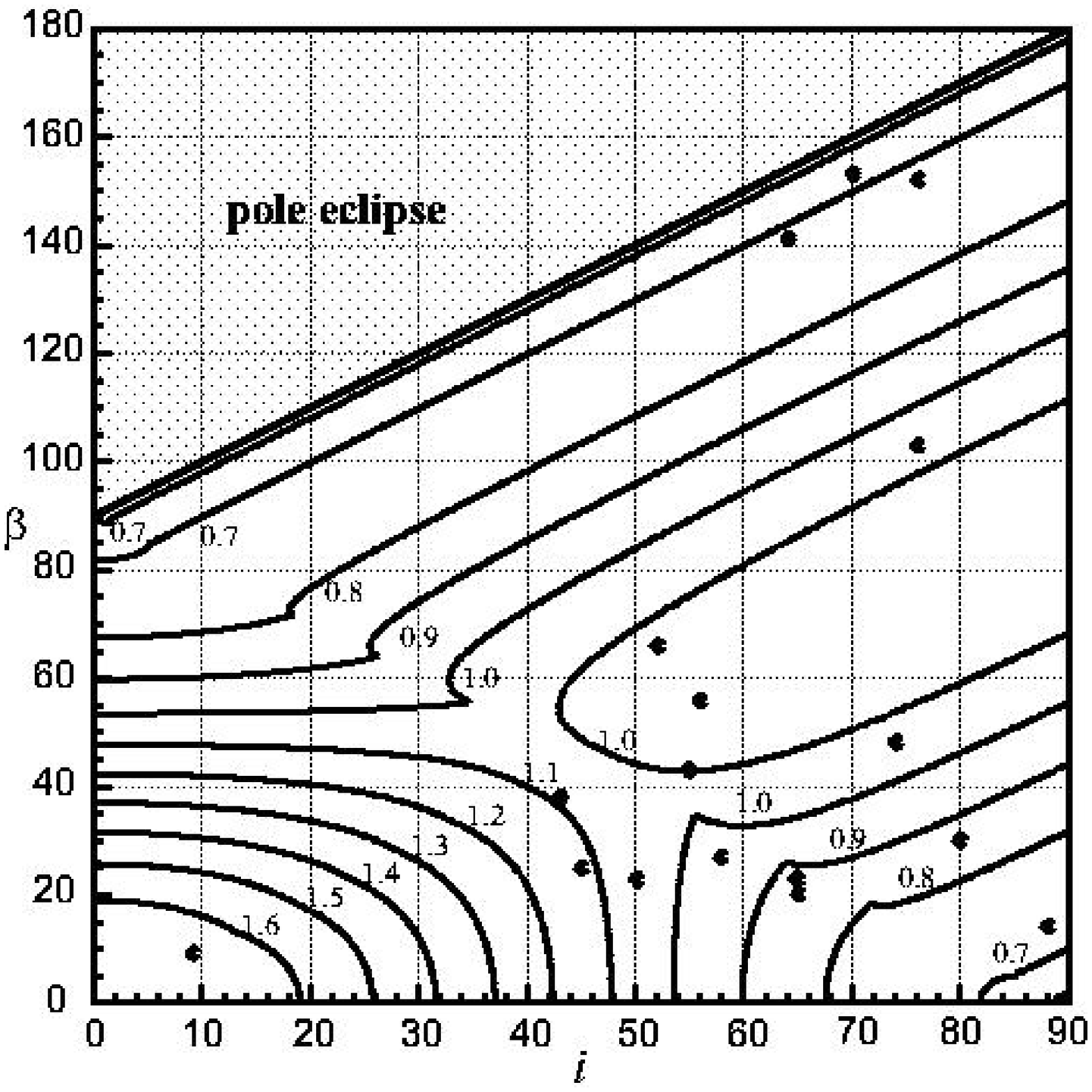}
\includegraphics[height=6.5cm]{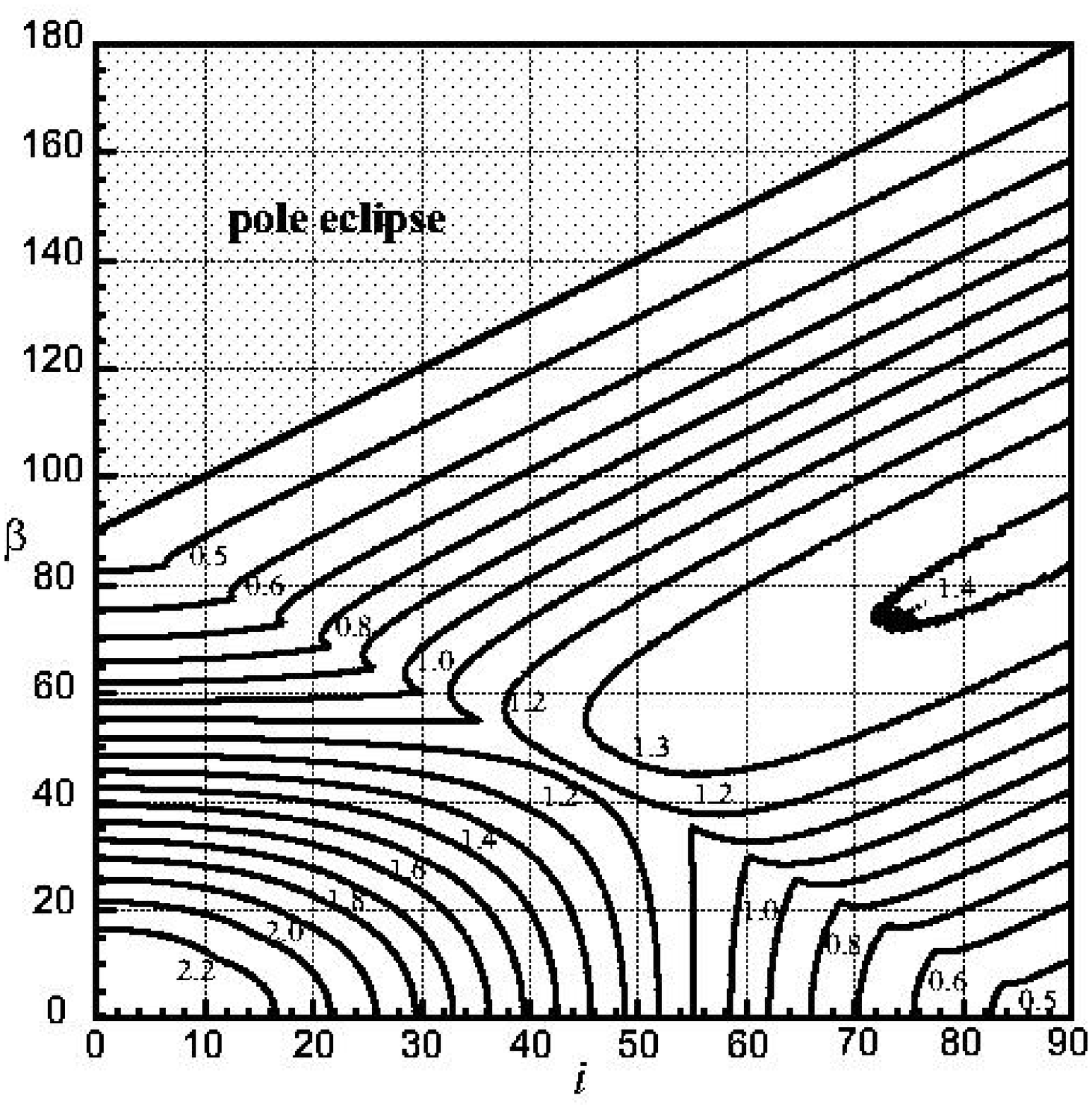}
\caption{The expected enhancement $\zeta$ to the average line intensity
calculated on the plane of ($i, \beta$),
assuming that accretion column exists only on one pole 
with the same collimation
as the case of V834 Cen (left) or the strong case shown in text (right).
Contour levels are shown in the figure.
The polars listed in table \ref{tbl:polar_geometry_list} are
also plotted in the left panel with tick marks.}
\label{fig:contour_i_beta_xi}
\end{figure*}

\begin{figure}
\includegraphics[height=5.6cm]{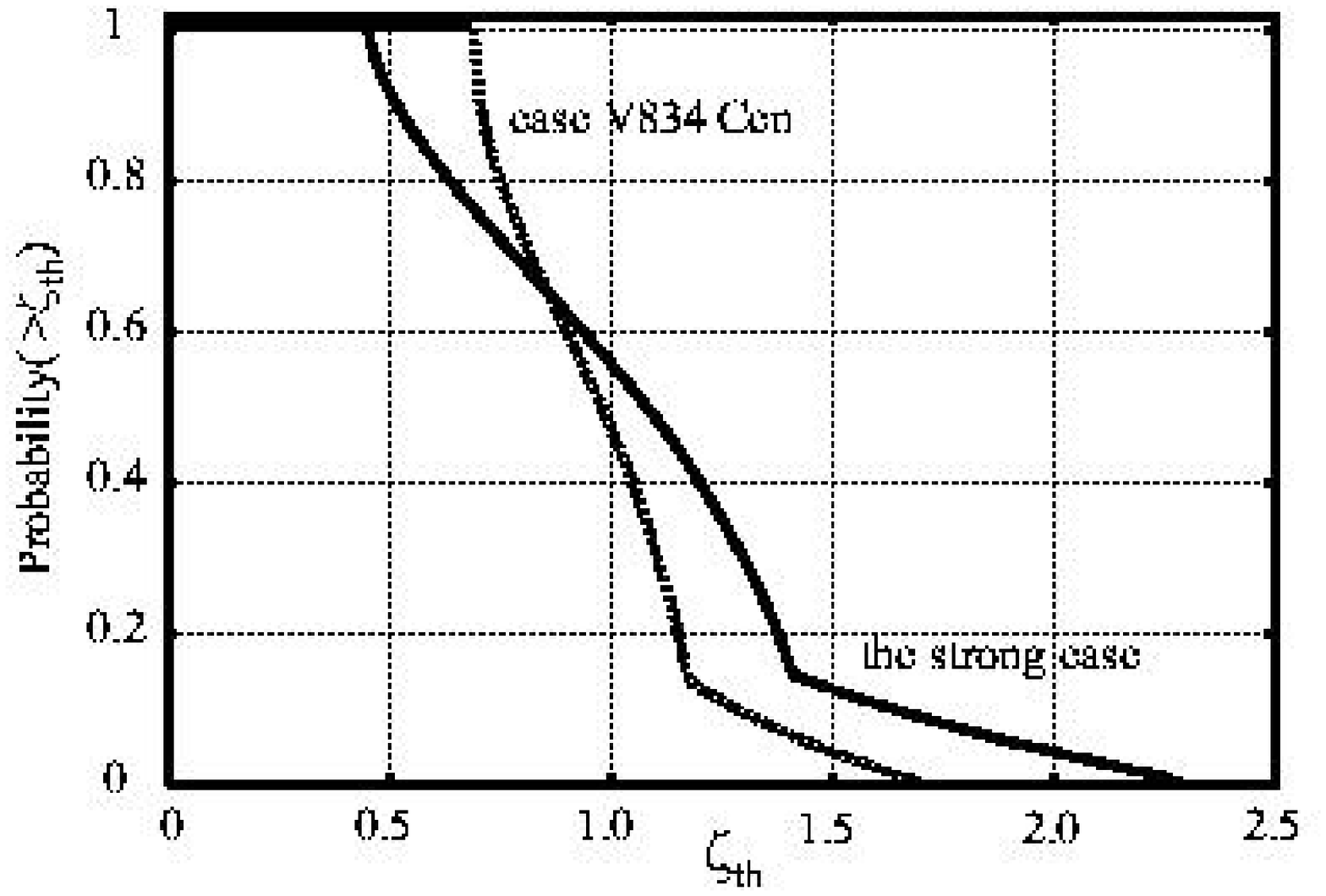}
\caption{The cumulative probability distribution of finding $\zeta$
higher than the specified threshold $\zeta_{\mbox{\tiny th}}$,
assuming that objects are randomly distributed in terms of $i$ and $\beta$.
The dotted and solid curves represent the case of V834 Cen
and the strong case, respectively.}
\label{fig:prob_i_beta_xi}
\end{figure}

In condition of $\zeta > 2$,
the extremely high face-value abundances of AX~J2315$-$0592
will be modified to be $\sim 1$ solar,
and abundance of RX~J1802.1$+$1804 or AX~J1842$-$0423
comes into the measured distribution of other polars within error.
We can expect a strong collimation with an adequate density
of $n_{\mbox{\tiny e}} \sim 10^{16}$ cm$^{-3}$
(figure \ref{fig:sim_result_various}b)
and an adequate shape of $h/r \sim 0.1 \mbox{--} 0.5$
(figure \ref{fig:sim_result_various}a) with typical radius $\sim 10^{7}$ cm.
Thus, one example of strong collimation calculated
for He-like iron resonance line is 
in the condition that temperature of 10keV, $r = 5 \times 10^{7}$ cm,
$n_{\mbox{\tiny e}} = 7.9 \times 10^{15}$ cm$^{-3}$ and
VEM = $10^{55}$ cm$^{-3}$, which yield an X-ray luminosity of 
$1.0 \times 10^{32}$ erg s$^{-1}$ (2 -- 10 keV).
Hereafter, we call this condition the strong case.
Figure \ref{fig:contour_i_beta_xi} shows 
the expected enhancement $\zeta$ of iron He-like line (sum of
resonance line, intercombination line and forbidden line)
with various geometrical conditions ($i$ and $\beta$).
In figure \ref{fig:prob_i_beta_xi},
we have converted the result of figure \ref{fig:contour_i_beta_xi}
into cumulative probability distribution of $\zeta$.
We hence expect $\zeta > 2$ at the condition of strong case
shown above, with a probability of $\sim$ 4.5 \%.
This estimate is in an rough agreement with the observation,
i.e. the three POLEs among the known $\sim$ 50 polars.
We reconfirms that the iron abundances derived from the observation
(Ezuka \& Ishida 1999) remains valid to within 30 \%
for a major fraction of polars.
Thus, we conclude that the mystery of POLEs have been solved by the proposed
line-collimation mechanism.

\section{Conclusion}
\label{section:conclusion}
In order to explain the extremely intense iron K lines
observed from several Galactic X-ray sources
(including two polars; section \ref{section:intro_fe}),
we have developed a scenario of POLEs,
that the iron lines from accretion poles of magnetic WDs are
axially collimated by the two mechanisms (section \ref{section:mechanism})
which become operational under large optical depths
for the resonant line scattering.
One is the geometrical effect
which becomes effective when the accretion column is rather short,
while the other is a physical mechanism
that the Doppler shifts
due to vertical velocity gradient in the post-shock flow
reduces the cross section for resonance scattering along the field lines.

In section \ref{section:monte_carlo},
we have carried out Monte-Carlo simulations,
and confirmed that the velocity-gradient effect,
augmented by the geometrical effect,
can enhance the iron K line EW in the pole-on direction
up to a factor of 3.0 as compared to the angular average.
This is higher than the maximum collimation available
with the geometrical effect alone
(Appendix \ref{section:geometrical_beaming}).

In order to experimentally verify our interpretation,
we have analyzed in section \ref{section:observation}
the ASCA data of V834 Cen, which has a suitable geometry in
that our line-of-sight to the accretion column changes
from $20^{\circ}$ to $70^{\circ}$ as the WD rotates.
Through detailed phase-resolved  X-ray spectroscopy,
the EW of the He-like iron-K$_\alpha$ line has been confirmed
to be enhanced by a factor of
$(\xi_{\mbox{\tiny pole}}/\xi_{\mbox{\tiny side}})^{\mbox{\tiny obs}}
= 1.87 \pm 0.54$. 
In section \ref{section:origin}, we have examined 
whether the observation can be explained away by any other mechanism,
and we have concluded that this observational result of V834 Cen
strongly reinforces our interpretation of POLEs.

Although the resonance lines are collimated with the proposed beaming effect,
the previous measurements of the distribution of metal abundances of polars
is considered still valid (section \ref{section:abundance})
except for POLEs.
With proposed mechanism,
the extremely high face-value abundances observed in POLEs
can be reconciled with the average abundance measured from the other polars.
Thus, the POLE scenario successfully solves the mystery of the extremely strong
iron lines observed from the three X-ray sources.

In addition, our scenario provides a new method of
unique determination of physical condition in the accretion column,
using the beaming factor of resonance lines
as a new observational information (section \ref{section:determine}).
This will be a powerful method for the next generation instruments.

\bigskip

We thank the members of the ASCA team
for spacecraft operation and data acquisition.

\appendix
\section[]{Plasma in an Accretion Column of a WD}
\label{section:accretion_column}

A flow of accreting matter captured by the magnetic field of WD 
channels into an accretion column, where a standing shock
forms and heats up the matter to a temperature $kT^{\mbox{\tiny sh}}$ of
\begin{eqnarray}
kT^{\mbox{\tiny sh}} &=& \frac{3}{8}
\frac{G M_{\mbox{\tiny WD}} \mu m_{\mbox{\tiny H}}}  {R_{\mbox{\tiny WD}}}
\nonumber \\
&=&
16 
\left( \frac{\mu}{0.615} \right)
\left( \frac{M_{\mbox{\tiny WD}}}{0.5 M_\odot} \right)
\left( \frac{R_{\mbox{\tiny WD}}}{10^9 \mbox{cm}}\right)^{-1}  \mbox{keV},
\label{eq:temparature}
\end{eqnarray}
where $k$ is the Boltzmann constant, $G$ is the gravitational constant,
$M_{\mbox{\tiny WD}}$ is the WD mass (typically $0.5 M_\odot$),
and $R_{\mbox{\tiny WD}}$ is the radius of the WD (typically $10^9$ cm).
So the plasma has a typical temperature of hard X-ray emitter,
forming an accretion column illustrated in figure \ref{fig:accretion_column}.
The velocity beneath the shock front $u^{\mbox{\tiny sh}}$
is described with relation to the free-fall velocity $u_{\mbox{\tiny ff}}$ as
\begin{eqnarray}
u^{\mbox{\tiny sh}} &=& \frac{u_{\mbox{\tiny ff}}}{4}
= \frac{1}{4}
\sqrt{\frac{2 G M_{\mbox{\tiny WD}}} {R_{\mbox{\tiny WD}}} }
\nonumber \\
&=& 0.9 \times 10^8 
\left( \frac{M_{\mbox{\tiny WD}}}{0.5 M_\odot}\right)^{1/2}
\left( \frac{R_{\mbox{\tiny WD}}}{10^9 \mbox{cm}}\right)^{-1/2}
\mbox{cm/s}.
\label{eq:velosity}
\end{eqnarray}
Assuming that the plasma is a single fluid
and that the abundance is one solar,
the electron density of the post-shock plasma
is given as 
\begin{eqnarray}
n_e^{\mbox{\tiny sh}}
&\simeq&
\left(
\frac{\dot{M}}
	{\pi r^2 u^{\mbox{\tiny sh}} \mu m_{\mbox{\tiny H}}}
\right)
\times 0.518
\nonumber \\
    &=&
7.7 \times 10^{15}
\left( \frac{\dot{M}}{10^{16} \mbox{g s}^{-1} } \right)
\left( \frac{M_{\mbox{\tiny WD}}}{0.5 M_\odot} \right)^{-1/2}
\left( \frac{R_{\mbox{\tiny WD}}}{10^9 \mbox{cm}} \right)^{1/2}
\nonumber \\ &\times&
\left( \frac{r}{5 \times 10^7 \mbox{cm}} \right)^{-2}
\mbox{cm}^{-3},
\label{equation:density}
\end{eqnarray}
where the value 0.518 is the fraction of electron density to the total density
assuming the solar abundances.
In the accretion column,
$kT$, $u$ and $n_e$ all have a vertical gradient
from the shock front toward the WD surface.
Numerically, the vertical profiles of these quantities
as a function of the distance $z$ from the WD surface,
normalized by $h$, are calculated by Aizu (1973) as 
\begin{equation}
\frac{kT}{kT^{\mbox{\tiny sh}}} = \frac{v}{u^{\mbox{\tiny sh}}}
= \left( \frac{n_e}{n_e^{\mbox{\tiny sh}}} \right)^{-1}
\simeq \left( \frac{z}{h} \right)^{\frac{2}{5}},
\label{eq:z_dependance}
\end{equation}
where each quantity is normalized to its value immediately
below the shock front;
$kT^{\mbox{\tiny sh}}$,
$u^{\mbox{\tiny sh}}$, and $n_e^{\mbox{\tiny sh}}$.

The column radius of the accretion column is typically $r = 5 \times 10^7$ cm.
Since the shock front is sustained by the pressure of the post-shock
plasma against the gravity, $h$ is described
by free-free cooling time scale $t_{\mbox{\tiny ff\_cool}}$
of the heated plasma
as $h \sim u^{\mbox{\tiny sh}} t_{\mbox{\tiny ff\_cool}}$.
According to Aizu (1973), $h$ is given more specifically as
\begin{eqnarray}
h&=&0.605 u^{\mbox{\tiny sh}} t_{\mbox{\tiny ff\_cool}}
\nonumber \\ 
  &=&1.9\times 10^7
   \left( \frac{kT^{\mbox{\tiny sh}}}{16 \mbox{keV}} \right) ^{\frac{1}{2}}
   \left( \frac{n_e^{\mbox{\tiny sh}}}
          {7.7\times 10^{15}\mbox{cm$^{-3}$}} \right) ^{-1}
\mbox{cm},
\label{eq:column_height}
\end{eqnarray}
where $t_{\mbox{\tiny ff\_cool}}$ is given by
\begin{eqnarray}
t_{\mbox{\tiny ff\_cool}} &\equiv&
\frac{3 n_e kT^{\mbox{\tiny sh}}}{2 \epsilon_{\mbox{\tiny ff}}}
\nonumber \\
&=& 0.35
\left( \frac{kT^{\mbox{\tiny sh}}}{16 \mbox{keV} }\right)^{\frac{1}{2}}
\left( \frac{n_e^{\mbox{\tiny sh}}}{7.7 \times 10^{15}
\mbox{cm $^{-3}$}}\right)^{-1}
\mbox{s}
\label{eq:ff_timescale}
\end{eqnarray}
with $\epsilon_{\mbox{\tiny ff}}$ being
the volume emissivity of free-free emission
[eq.\ (5.15) in Rybicki \& Lightman 1979].
Note that
the ion to electron energy transfer time scale $t_{\mbox{\tiny eq}}$
\begin{eqnarray}
t_{\mbox{\tiny eq}} &=& 5.6 \times 10^{-3}
\left( \frac{kT^{\mbox{\tiny sh}}}{16 \mbox{keV} }\right)^{\frac{3}{2}}
\left( \frac{n_e^{\mbox{\tiny sh}}}{7.7 \times 10^{15}
\mbox{cm $^{-3}$}}\right)^{-1}
\mbox{s}
\label{eq:eq_timescale}
\end{eqnarray} 
[see eq (5.31) in Spitzer 1962]
is much shorter than $t_{\mbox{\tiny ff\_cool}}$, 
so the ions and electrons are thought to share the same temperature.

At the temperature of a few tens keV [equation (\ref{eq:temparature})],
the electron scattering dominates the opacity in the hard X-ray band.
Actually, the optical depth of the accretion column for
free-free absorption, $\tau_{\mbox{\tiny ff}}$, is given,
relative to the electron scattering optical depth
$\tau_{\mbox{\tiny T}}$ [equation (\ref{eq:thomson_opt})] as
\begin{eqnarray}
\tau_{\mbox{\tiny ff}} &=& 0.88 \times 10^{-10}
\tau_{\mbox{\tiny T}} 
\left( \frac{E}{6 \mbox{keV}} \right)^{-2}
\left( \frac{kT}{16 \mbox{keV}} \right)^{-\frac{3}{2}}
\nonumber \\ && \times 
\left( \frac{n_e^{\mbox{\tiny sh}}}
            {7.7 \times 10^{15}\mbox{cm$^{-3}$}} \right)
\label{eq:opt_ff}
\end{eqnarray}
[see equation (5.18) in Rybicki \& Lightman 1979]. 
Thus, the free-free absorption is negligible compared to
Thomson scattering.

The cross section of resonance scattering $\sigma_{\mbox{\tiny RS}}$
for a photon with energy $E_{\mbox{\tiny 0}}$ can be described generally as
\begin{equation}
\sigma_{\mbox{\tiny RS}}
= \frac{\pi e^2}{m_{\mbox{\tiny i}}c} f_{12}
\frac{1}{\sqrt{2\pi} \Delta E}
\exp\left\{- \frac{(E_{\mbox{\tiny 0}} - E_{\mbox{\tiny RS}})^2}
{2 \Delta E^2} \right\}
\mbox{cm$^{-3}$}
\label{eq:reso_cross}
\end{equation}
where $f_{12}$ is the oscillator strength for Lyman-$\alpha$ transition
(n = 1 to 2),
$E_{\mbox{\tiny RS}}$ is the resonance energy in the rest frame,
and $\Delta E$ is a resonance energy width
[eq.\ (10.70) in Rybicki \& Lightman 1979].
Numerically, the first factor is
$\frac{\pi e^2}{m_{\mbox{\tiny Fe}}c} f_{12} =1.2\times10^{-17}$ cm$^{-3}$,
and the energy width $\Delta E$ in the second factor of Gaussian is
determined by equations (\ref{eq:v_therm}) and (\ref{eq:thermal_deltaE}).
Then, the cross section of resonance scattering at the line-center energy
is given as 
\begin{equation}
\sigma_{\mbox{\tiny RS}} = 2.0 \times 10^{-18}
\left(  \frac{E_{\mbox{\tiny 0}} }{ 6.695 \mbox{keV} }  \right)
\left( \frac{kT}{16 \mbox{keV}} \right)^{-1/2}  \mbox{cm$^{2}$}.
\label{eq:reso_cross_sec}
\end{equation}

\section[]{Geometrical beaming in the accretion column}
\label{section:geometrical_beaming}

\begin{figure}
\includegraphics[height=6cm]{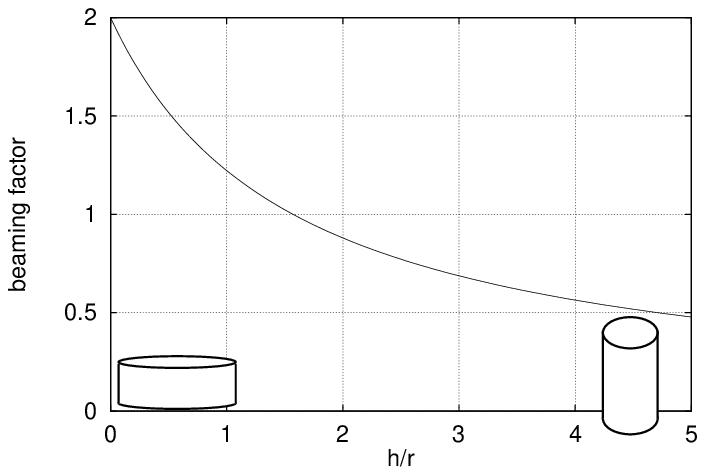}
\caption{The collimation factor of geometrical beaming.
The flux at $\theta = 0$ to the averaged flux
[equation (\ref{eq:geometrical_beaming_mean})].}
\label{fig:geometrical_beaming}
\end{figure}

How much enhancement can we expect in ``geometrical beaming''
in the accretion column
(see $\S$\ref{section:mechanism_geometrical_beaming})?
Consider the case that
the resonance line photons can only escape from the surface of the column.
The directional photon flux emerging from the column is
given as
\begin{equation}
f(\theta) d \cos\theta \propto \pi r^2\cos\theta + 2rh\sin\theta,
\label{eq:geometrical_profile}
\end{equation}
where $\theta$ is the angle measured from the column axis.
Therefore, the flux along $\theta \simeq 0$ is enhanced by a factor
\begin{equation}
\frac{f(\theta)}{<f>} = \frac{\pi r^2}{ \frac{1}{2} \pi r^2 + \frac{\pi}{2}rh}
= \frac{2}{1 + \frac{2}{\pi}\left(\frac{h}{r}\right)},
\label{eq:geometrical_beaming_mean}
\end{equation}
where $<f>$ is the average of $f(\theta)$ over $\theta$.
At the coin-shaped limit $(\frac{h}{r}\rightarrow 0)$,
this factor approaches $2$ (figure \ref{fig:geometrical_beaming}).

\label{lastpage}

\begin{thebibliography}{}


\bibitem{aizu}
Aizu, K. 1973, Prog. Theor. Phys., 49, 1184

\bibitem{v834_optical}
Bailey, J., Axon, D.\ J., Hough, J.\ H., Watts, D.\ J.,
Giles, A.\ B., Greenhill, J.\ G.
1983, MNRAS, 205, 1

\bibitem{geom_WWHor}
Bailey, J., Wickramasinghe, D.\ T., Hough, J.\ H., Cropper, M.
1988, MNRAS, 234, 19

\bibitem{reflect1}
Beardmore, A.\ P., Ramsay, G., Osborne, J.\ P.,
Mason, K.\ O., Nousek, J.\ A., Baluta, C.
1995, MNRAS, 273, 742

\bibitem{geom_VYFor}
Beuermann, K., Thomas, H.\ -., Giommi, P., Tagliaferri, G.
and Schwope, A.\ D.
1989, A\&A, 219, 7

\bibitem{sis1}
Burke, B.E., Mountain, R.W., Daniels, P.J., Cooper, M.J., Dolat, V.S.
1994, IEEE Trans. Nucl. Sci., NS-41, 375

\bibitem{geom_J1015}
Burwitz, V., Reinsch, K., Schwope, A.\ D., Hakala, P.\ J., 
Beuermann, K., Rousseau, T., Thomas, H.\ -., 
Gansicke, B.\ T., Piirola, V., Vilhu, O.
1998, A\&A, 331, 262

\bibitem{geom_QQVul2}
Catal{\'a}n, M.\ S. and Schwope, A.\ D. and Smith, R.\ C.
1999, MNRAS, 310, 123

\bibitem{cropper}
Cropper, M. 1990, SSRv, 54, 195

\bibitem{reflect2}
Done, C., Osborne, J.\ P., Beardmore, A.\ P.
1995, MNRAS, 276, 483

\bibitem{sscyg}
Done, C. Osborne, J.\ P.
1997, MNRAS, 288, 649

\bibitem{reflect3}
Done, C. Magdziarz, P.
1998, MNRAS, 298, 737

\bibitem{paper2}
Ezuka, H. Ishida, M.
1999, ApJS, 120, 277

\bibitem{geom_UZFor}
Ferrario, L., Wickramasinghe, D.\ T., Bailey, J., 
Tuohy, I.\ R., Hough, J.\ H.
1989, ApJ, 337, 832

\bibitem{exhya}
Fujimoto, R. Ishida, M. 1997, ApJ, 474, 774

\bibitem{rxj1802_rosat}
Greiner, J., Remillard, R.\ A., Motch, C.
1998, A\&A, 336, 191

\bibitem{geom_V1309ORI}
Harrop-Allin, M.\ K., Cropper, M., Potter, S.\ B., 
Dhillon, V.\ S., Howell, S.\ B.
1997, MNRAS, 288, 1033

\bibitem{hoshi}
Hoshi, R. 1973, Prog. Theor. Phys., 49, 776

\bibitem{ishida_91}
Ishida, M. 1991, Ph.D. thesis. Univ. of Tokyo

\bibitem{amher_ishida}
Ishida, M., Matsuzaki, K., Fujimoto, R., Mukai, K., Osborne, J.\ P., 
1997, MNRAS, 287, 651

\bibitem{rxj1802_asca}
Ishida, M., Greiner, J., Remillard, R.\ A., Motch, C.,
1998, A\&A, 336, 200

\bibitem{gis2}
Makishima, K.,  Tashiro, M.,  Ebisawa, K.,  Ezawa, H.,  
Fukazawa, Y.,  Gunji, S.,  Hirayama, M.,  Idesawa, E.,  
Ikebe, Y.,  Ishida, M.,  Ishisaki, Y.,  Iyomoto, N.,  
Kamae, T.,  Kaneda, H.,  Kikuchi, K.,  Kohmura, Y.,  
Kubo, H.,  Matsushita, K.,  Matsuzaki, K.,  Mihara, T.,  
Nakagawa, K.,  Ohashi, T.,  Saito, Y.,  Sekimoto, Y.,  
Takahashi, T.,  Tamura, T.,  Tsuru, T.,  Ueda, Y., Yamasaki, N.\ Y.
1996, PASJ, 48, 171

\bibitem{mew}
Mewe, R., Gronenschild, E.\ H.\ B.\ M., van den Oord, G.\ H.\ J.
1985, A\&AS, 62, 197

\bibitem{axj2315_misaki}
Misaki, K., Terashima, Y., Kamata, Y., Ishida, M., Kunieda, H.,
Tawara, Y. 1996, ApJ, 470, 53

\bibitem{norton}
Norton, A.\ J. Watson, M.\ G.
1989, MNRAS, 237, 853

\bibitem{gis1}
Ohashi, T.,  Ebisawa, K.,  Fukazawa, Y.,  Hiyoshi, K.,  
Horii, M.,  Ikebe, Y.,  Ikeda, H.,  Inoue, H.,  
Ishida, M.,  Ishisaki, Y.,  Ishizuka, T.,  Kamijo, S.,  
Kaneda, H.,  Kohmura, Y.,  Makishima, K.,  Mihara, T.,  
Tashiro, M.,  Murakami, T.,  Shoumura, R.,  Tanaka, Y.,  
Ueda, Y.,  Taguchi, K.,  Tsuru, T., Takeshima, T.
1996, PASJ, 48, 157O

\bibitem{rybicki}
Rybicki,G.B.\& Lightman,A.P. 1979, Radiative Processes in Astrophysics
(New Yorkl Wiley)

\bibitem{geom_QQVul1}
Schwope, A.\ D., Catal\'an, M.\ S., Beuermann, K.,
Metzner, A.\ ;, Smith, R.\ C., Steeghs, D.
2000, MNRAS, 313, 533

\bibitem{spitzer}
Spitzer, L. 1962, Physics of Fully Ionized Gases (New Yorkl Wiley)

\bibitem{paper1}
Terada, Y., Kaneda, H., Makishima, K., Ishida, M.,
Matsuzaki, K., Nagase, F. Kotani, T.
1999, PASJ, 51, 39

\bibitem{axj2315_opt}
Thomas, H.\ -., Reinsch, K.
1996, A\&A, 315L, 1

\bibitem{v834_freq}
Schwope, A.\ D., Thomas, H.\ -., Beuermann, K., Reinsch, K.
1993, A\&A, 267, 103

\bibitem{xrt}
Serlemitsos, P.\ J., Jalota, L., Soong, Y., Kunieda, H., Tawara, Y.,
Tsusaka, Y., Suzuki, H., Sakima, Y., Yamazaki, T., Yoshioka, H.,
Furuzawa, A., Yamashita, K. Awaki, H., Itoh, M., Ogasaka, Y.,
Honda, H., Uchibori, Y.
1995, PASJ, 47, 105

\bibitem{geom_ARUMa}
Szkody, P., Vennes, S.\ ;., Schmidt, G.\ D., 
Wagner, R.\ M., Fried, R., Shafter, A.\ W., Fierce, E.
1999, ApJ, 520, 841

\bibitem{geom_AMHER}
Wickramasinghe, D.\ T., Bailey, J., Meggitt, S.\ M.\ A.,
Ferrario, L., Hough, J., Tuohy, I.\ R.
1991, MNRAS, 251, 28

\bibitem{wu_mass}
Wu, K., Chanmugam, G., Shaviv, G.
1995, ApJ, 455, 260

\bibitem{sis2}
Yamashita, A., Dotani, T., Bautz, M., Crew, G., Ezuka, H.,
Gendreau, K., Kotani, T., Mitsuda, K., Otani, C., Rasmussen, A.,
Ricker, G., Tsunemi, H.
1997, IEEE Trans. Nucl. Sci., NS-44, 847

\end{thebibliography}
\end{document}